\documentclass{vldb}

\usepackage[english]{babel}
\usepackage[labelfont=bf]{caption}
\usepackage{subcaption}
\usepackage{amsmath}
\usepackage{amsfonts}
\usepackage{amssymb}
\usepackage{cite}
\usepackage{graphicx}
\usepackage{thm-restate}
\usepackage[usenames,dvipsnames,table]{xcolor}
\usepackage{algpseudocode}
\usepackage{algorithm}
\usepackage{balance}
\usepackage{xspace}
\usepackage[capitalize,noabbrev]{cleveref}
\usepackage{outlines}
\usepackage{outlines}
\usepackage{multicol}
\usepackage{multirow}
\usepackage{bm}
\usepackage{array}
\usepackage{tabularx}
\usepackage{dcolumn}
	\newcolumntype{d}[1]{D{.}{.}{#1|}}
\usepackage{booktabs}
\usepackage{color,soul}
\usepackage[group-separator={,}]{siunitx}

\begin{document}


\newtheorem{remark}{Remark}
\newtheorem{theorem}{Theorem}
\newtheorem{lemma}{Lemma}
\newtheorem{definition}{Definition}
\newtheorem{problem}{Problem}
\newtheorem{proposition}{Proposition}
\newtheorem{corollary}{Corollary}
\newtheorem{example}{Example}
\newtheorem{conjecture}{Conjecture}
\newtheorem{operator}{Operator}
\def\map{{\mbox{\sf MAP}}}
\def\plus{{+}}
\def\opt{{\mbox{\sf OPT}}}
\def\metaopt{\ensuremath{{\mbox{\sf OPT}_{\sf HDMM}}}}
\def\WW{\mathbb{W}}
\def\Wlog{\mathcal{W}}
\def\AA{\mathbb{A}}
\def\BB{\mathbb{B}}
\def\CC{\mathbb{C}}
\def\GG{\mathbb{G}}
\def\MM{\mathbb{M}}
\def\QQ{\mathbb{Q}}
\def\optgp{\mbox{\sf OPT}_{0}}
\def\optk{\mbox{\sf OPT}_{\otimes}}
\def\optkk{\mbox{\sf OPT}_{+}}
\def\optm{\mbox{\sf OPT}_{\mathsf{M}}}

\def\Imp{\mbox{\sf ImpVec}}
\def\mult{\mbox{\sf Multiply}}
\def\invert{\mbox{\sf LstSqr}}

\newcommand{\revision}[1]{#1}

\newcommand{\tighttimes}{\!\times\!}

\newcommand{\algoname}[1]{{\textsc{#1}}\xspace}
\newcommand{\sys}{\algoname{HDMM}}
\newcommand{\sysFull}{High-Dimensional Matrix Mechanism\xspace}
\newcommand{\LMW}{\algoname{LM}}
\newcommand{\Identity}{\algoname{Identity}}
\newcommand{\Workload}{\ensuremath{MM(W)}}
\newcommand{\PrivBayes}{\algoname{PrivBayes}}
\newcommand{\matrixMech}{\algoname{MM}}
\newcommand{\HB}{\algoname{HB}}
\newcommand{\Privelet}{\algoname{Privelet}}
\newcommand{\Quadtree}{\algoname{Quadtree}}
\newcommand{\GreedyH}{\algoname{GreedyH}}
\newcommand{\DataCube}{\algoname{DataCube}}
\newcommand{\DAWA}{\algoname{DAWA}}
\newcommand{\LRM}{\algoname{LRM}}

\newcommand{\datasetname}[1]{{\emph{#1}}\xspace}
\newcommand\Census{\datasetname{CPH}}
\newcommand\Person{\datasetname{Person}}
\newcommand\Adult{\datasetname{Adult}}
\newcommand\CPS{\datasetname{CPS}}
\newcommand\Patent{\datasetname{Patent}}
\newcommand\Taxi{\datasetname{Taxi}}

\newcommand{\workloadname}[1]{{\emph{#1}}\xspace}
\newcommand\FixedWidthRange{\workloadname{Width 32 Range}}
\newcommand\PrefixOneD{\workloadname{Prefix 1D}}
\newcommand\PrefixTwoD{\workloadname{Prefix 2D}}
\newcommand\PrefixThreeD{\workloadname{Prefix 3D}}
\newcommand\PermutedRange{\workloadname{Permuted Range}}
\newcommand\PrefixIdentity{\workloadname{Prefix Identity}}
\newcommand\SFOne{\workloadname{SF1}}
\newcommand\SFOnePlus{\workloadname{SF1+}}
\newcommand\AllMarginals{\workloadname{All Marginals}}
\newcommand\TwoWayMarginals{\workloadname{2-way Marginals}}
\newcommand\ThreeWayMarginals{\workloadname{3-way Marginals}}
\newcommand\AllRangeMarginals{\workloadname{All Range-Marginals}}
\newcommand\TwoWayRangeMarginals{\workloadname{2-way Range-Marginals}}
\newcommand\ThreeWayRange{\workloadname{All 3-way Ranges}}

\newcommand{\errRatio}{\ensuremath{Ratio(W, \mathcal{A}_{other})}\xspace}

\newcommand{\NA}{-}
\newcommand{\NS}{*}


\newcommand{\op}[1]{{\sc #1}\xspace}
\newcommand{\stitle}[1]{\vspace{-2.5mm}\paragraph*{#1}}
\newcommand{\eat}[1]{}
\newcommand{\eatrev}[1]{}
\newcommand{\mybullet}{\vspace{1mm}\noindent$\bullet$~~}

\newcommand{\todo}[1]{[[\emph{\color{teal}TODO: #1}]]}
\newcommand{\gm}[1]{[[\emph{\color{red}GM: #1}]]}
\newcommand{\ry}[1]{[[\emph{\color{blue}RM: #1}]]}
\newcommand{\mh}[1]{[[\emph{\color{red}MH: #1}]]}
\newcommand{\am}[1]{[[\emph{\color{magenta}AM: #1}]]}

\newcommand{\gmref}[2]{{\color{cyan} #1}~[[\emph{{\color{red} GM: #2}}]]}
\newcommand{\ryref}[2]{{\color{red} #1}~[[\emph{{\color{blue} RM: #2}}]]}
\newcommand{\mhref}[2]{{\color{cyan} #1}~[[\emph{{\color{red} MH: #2}}]]}
\newcommand{\amref}[2]{{\color{cyan} #1}~[[\emph{{\color{magenta} AM: #2}}]]}

\renewcommand*\Call[2]{\textproc{#1}(#2)}



\def\Lap{\mbox{Lap}}
\def\sens{\Delta}
\def\LM{\mbox{LM}}

\newcommand{\Lone}[1]{\left\Vert #1  \right\Vert_1}
\newcommand{\norm}[1]{\left\lVert#1\right\rVert}
\newcommand{\set}[1]{\{#1\}}   

\newcommand{\vect}[1]{\bm{#1}}
\newcommand{\matr}[1]{\bm{#1}}
\def\W{{\matr{W}}}
\def\A{\matr{A}}
\def\B{\matr{B}}
\def\C{\matr{C}}
\def\D{\matr{D}}
\def\L{\matr{L}}
\def\I{\matr{I}}
\def\T{\matr{T}}
\def\R{\matr{R}}
\def\P{\matr{P}}
\def\S{\matr{S}}
\def\M{\matr{M}}
\def\Y{\matr{Y}}
\def\X{\matr{X}}
\def\x{\vect{x}}
\def\q{\vect{q}}
\def\a{\vect{a}}
\def\u{\vect{u}}
\def\v{\vect{v}}
\def\w{\vect{w}}

\def\xhat{\bar{\x}}
\def\y{\vect{y}}

\newcommand{\Wnt}{\ensuremath{\mathcal{W}_{\mbox{\tiny SF1}}}}
\newcommand{\Wst}{\ensuremath{\mathcal{W}_{\mbox{\tiny SF1+}}}}

 \def\db{I}  
 \def\nbrs{nbrs}

 \def\algG{\mathcal{K}}  

\vldbTitle{Optimizing error of high-dimensional statistical queries under differential privacy}
\vldbAuthors{Ryan McKenna, Gerome Miklau, Michael Hay, Ashwin Machan\-avajjhala}
\vldbDOI{https://doi.org/10.14778/3231751.3231769}

\title{Optimizing error of high-dimensional statistical queries under differential privacy\titlenote{Views expressed in this paper are those of  authors and do not necessarily reflect the views of the U.S. Census Bureau.}}

\numberofauthors{1}
\author{
	\alignauthor Ryan McKenna\textsuperscript{\textdagger},
	Gerome Miklau\textsuperscript{\textdagger,\textasteriskcentered},
	Michael Hay\textsuperscript{$\ddagger$,\textasteriskcentered},
	Ashwin Machanavajjhala\textsuperscript{\textasteriskcentered\textasteriskcentered}\\
	\affaddr{\makebox[0pt][r]{\textsuperscript{\textdagger}}
			Univ. of Massachusetts, Amherst, College of Information and Computing Sciences}\\
	\affaddr{\makebox[0pt][r]{\textsuperscript{$\ddagger$}}
		Colgate University, Dept. of Computer Science}\\
	\affaddr{\makebox[0pt][r]{\textsuperscript{\textasteriskcentered\textasteriskcentered}}
		Duke University, Dept. of Computer Science}\\
	\affaddr{\makebox[0pt][r]{\textsuperscript{\textasteriskcentered}}
			U.S. Census Bureau, Suitland, MD}\\
	\affaddr{\{rmckenna, miklau\}@cs.umass.edu  \ \ \ mhay@colgate.edu  \ \ \ ashwin@cs.duke.edu}
}

\maketitle


\pagestyle{plain}
\pagenumbering{arabic}

\begin{abstract}
Differentially private algorithms for answering sets of predicate counting queries on a sensitive database have many applications. Organizations that collect individual-level data, such as statistical agencies and medical institutions, use them to safely release summary tabulations. However, existing techniques are accurate only on a narrow class of query workloads, or are extremely slow, especially when analyzing more than one or two dimensions of the data.

In this work we propose \sys, a new differentially private algorithm for answering a workload of predicate counting queries, 
that is especially effective for higher-dimensional datasets.  
\sys represents query workloads using an implicit matrix representation and exploits this compact representation to efficiently search (a subset of) the space of differentially private algorithms for one that answers the input query workload with high accuracy. We empirically show that \sys can efficiently answer queries with lower error than state-of-the-art techniques on a variety of low and high dimensional datasets.
\end{abstract}

\section{Introduction}  \label{sec:introduction}
\begin{table*}
	\caption{\label{fig:alg_overview} Overview of the {\em High Dimensional Matrix Mechanism} (\sys), compared with the Matrix Mechanism (\matrixMech)~\cite{li2010optimizing}.}
\centering
{\small
\subcaptionbox{\label{fig:mm} The Matrix Mechanism (\matrixMech)~\cite{li2010optimizing}}{
		\begin{tabular}{rlcl}
			\multicolumn{4}{l}{\textbf{Input:} workload $\W$, in matrix form}\\
			\multicolumn{4}{l}{\hspace{.45in} data $\x$, in vector form}\\
			\multicolumn{4}{l}{\hspace{.45in} privacy parameter $\epsilon$}\\
			\hline \hline
			& \\ 
 			{\sf SELECT} $\begin{cases}\end{cases}$ \hspace{1ex} \qquad  & $\A$ & = & $\opt_{MM}(\W)$ \\
			\multirow{2}{*}{{\sf MEASURE} $\begin{cases} \mathstrut \end{cases}$} &$\a$ & = & $\A\x$ \\
			& $\y$ & = & $\a + Lap(||\A||_1 / \epsilon)$ \\
			\multirow{2}{*}{{\sf RECONSTRUCT} $\begin{cases} \mathstrut \end{cases}$} & $\xhat$ & = & $\A^+\y$ \\
			& $ans$ & = & $\W \xhat$
		\end{tabular}
}%
\hfill
\subcaptionbox{HDMM Overview}{
		\begin{tabular}{lcll}
			\textbf{Input:} & \multicolumn{3}{l}{workload $\Wlog$, in logical form}\\
			& \multicolumn{3}{l}{data $\x$, in vector form}\\
			& \multicolumn{3}{l}{privacy parameter $\epsilon$}\\
			\hline \hline
			 $\WW$ & = & $\Imp(\Wlog)$ & \emph{// Compact vector representation}\\
			  $\AA$ & = & $\metaopt(\WW)$ & \emph{// Optimized strategy selection}\\
			 $\a$ & = & $\mult(\AA, \x)$ & \emph{// Strategy query answering} \\
			$\y$ & = & $\a + Lap(||\AA||_1 / \epsilon)$ & \emph{// Noise addition}  \\
			$\xhat$ & = & $\invert(\AA, \y)$  & \emph{// Inference}\\
			$ans$ & = & $\mult(\WW, \xhat)$ & \emph{// Workload answering}
		\end{tabular}
} }
\vspace{1ex}
\end{table*}

Institutions like the U.S. Census Bureau and Medicare regularly release summary statistics about individuals, including population statistics cross-tabulated by demographic attributes \cite{census-sf1,onthemap} and tables reporting on hospital discharges organized by medical condition and patient characteristics \cite{hcupnet}.
%
%
These data have the potential to reveal sensitive information, especially through joint analysis of multiple releases~\cite{onthemap:icde08,sigmod:haney17,vaidya2013hcupnet}.  Differential privacy~\cite{dwork2006calibrating,Dwork14Algorithmic} has become the dominant standard for ensuring the privacy of such data releases.  An algorithm for releasing statistics over a dataset satisfies
$\epsilon$-differential privacy if adding or removing a single record in the input dataset does not result in a significant change in the output of the algorithm. The allowable change is determined by $\epsilon$, also called the privacy-loss budget. If each record in the input corresponds to a unique individual, this notion gives a compelling privacy guarantee \cite{tods:Kifer14}.

We consider the problem of releasing answers to a \emph{workload} (i.e., a set) of \emph{predicate counting queries} while satisfying $\epsilon$-differential privacy.  Predicate counting queries have the form \texttt{SELECT Count(*) FROM R WHERE $\phi$}, where $\phi$ is any boolean formula over the attributes in \texttt{R}.
(This problem formulation also supports group-by queries, each of which can be rewritten into a set of predicate counting queries, one query per possible group.)
Workloads of such queries are quite versatile, expressing histograms, multi-dimensional range queries, data cubes, marginals, or arbitrary combinations thereof.

There has been a plethora of work on differentially private techniques for answering sets of queries including work establishing theoretical lower bounds~\cite{bhaskara2012unconditional, hardt2010geometry,nikolov2013geometry} and practical  algorithms~\cite{zhang16privtree,yuan2016convex,li2015matrix,zhangtowards,xiao2014dpcube,qardaji2014priview,li2014data,Yaroslavtsev13Accurate,xu2013differential,qardaji2013understanding,qardaji2013differentially,yuan2012low,xu12histogram,li2012adaptive,cormode2012differentially,Acs2012compression,xiao2011differential,ding2011differentially,li2010optimizing,hay2010boosting,barak2007privacy,qardaji2014priview,Zhang2014}.

One class of techniques answers the queries of interest on the database and then uses the Laplace Mechanism to add noise, calibrated to their \emph{sensitivity}, or the maximum change in answers resulting from one change in the input database \cite{dwork2006calibrating,xiao2008output,ebadi2015personal,inan2016graph}. These techniques can answer queries using off-the-shelf systems (queries in SQL and data in relational form), and thus can be implemented efficiently \cite{mcsherry2009pinq,johnson2017elastic}.
However, a key limitation of this class is that because the noise is calibrated on a per-query basis, they fail to exploit workload structure and thus add more noise than is strictly necessary, resulting in suboptimal accuracy.

A second, more sophisticated, approach to query answering generalizes the Laplace Mechanism by first \textbf{selecting} a new set of \emph{strategy} queries, then \textbf{measuring} the strategy queries using the Laplace mechanism, and \textbf{reconstructing} answers to the input queries from the noisy measurements.  Choosing an effective query answering strategy (different from the workload) can result in orders-of-magnitude lower error than the Laplace mechanism, with no cost to privacy.

An example of a technique from the select-measure-re\-con\-struct paradigm is the Matrix Mechanism (\matrixMech)~\cite{li2015matrix}, illustrated in \cref{fig:mm}.
The \matrixMech, and other techniques in this paradigm, represent the database and queries in vector form, expressed over the full domain of each tuple (the product of the domains of the attributes). The vector representation allows these techniques to compute the sensitivity of sets of queries using a matrix norm, and to use inference algorithms based on linear algebra to reconstruct answers from noisy measurements.  In this vector form, the selection step corresponds to selecting a query matrix $\A$ (the strategy), and the measurement step reduces to computing the matrix-vector product between $\A$ and the data vector $\x$.

Many recent algorithms fall within the select-measure-reconstruct paradigm~\cite{zhang16privtree,yuan2016convex,li2015matrix,zhangtowards,xiao2014dpcube,qardaji2014priview,li2014data,Yaroslavtsev13Accurate,xu2013differential,qardaji2013understanding,qardaji2013differentially,yuan2012low,xu12histogram,li2012adaptive,cormode2012differentially,Acs2012compression,xiao2011differential,ding2011differentially,li2010optimizing,hay2010boosting,li2015matrix}, differing primarily in the measurement selection step.
We can characterize measurement selection as a search problem over a space of strategies, distinguishing prior work in terms of key algorithmic design choices: the search space, the cost function, and the type of search algorithm (greedy, local, global, etc.).
%
These design choices impact the three key performance considerations: accuracy, runtime, and scalability (in terms of increasing dataset dimensionality).

At one extreme are techniques that explore a narrow search space, making them efficient and scalable but not particularly accurate (in particular, their search space may include accurate strategies only for a limited class of workloads).  For example, \HB~\cite{qardaji2013understanding} considers strategies consisting of hierarchically structured interval queries.  It performs a simple search to find the branching factor of the hierarchical strategy that minimizes an error measure that assumes the workload consists of all range queries (regardless of the actual input workload).  It is efficient and can scale to higher dimensions, but it achieves competitive accuracy only when the workload consists of range queries and the data is low dimensional.

At the other extreme are techniques that search a large space, and adapt to the workload by finding a strategy within that space that offers low error on the workload, thereby making them capable of producing a more accurate strategy for the particular input. 
However, this increased accuracy comes at the cost of high runtime and poor scalability.  This is exemplified by \matrixMech, which solves a rank-constrained semi-definite program to find the \emph{optimal} solution.  Unfortunately, the optimization program is infeasible to execute on any non-trivial input workload.

In short, there is no prior work that is accurate for a wide range of input workloads, sufficiently fast, and capable of scaling to large multi-dimensional domains.

\paragraph*{Overview of approach and contributions} \label{sec:sub:overview}

We describe the \sysFull (\sys), a new algorithm for answering workloads of predicate counting que\-ries.  
While similar in spirit to the matrix mechanism, there are a number of innovations that make it more efficient and scalable.  We contrast the two algorithms in \cref{fig:alg_overview}.

First, \matrixMech represents query workloads as fully-mater\-ial\-ized matrices, while \sys uses a compact \emph{implicit} matrix representation of the logical queries, which we call a \emph{union of products} (\cref{sec:implicit}), for which query sizes are not exponential in the number of attributes.  In the use case we will describe soon, the matrix representation of one of the workloads would be 22TB; in contrast, our most compact representation of this workload is just 687KB.  
Without this innovation it is infeasible merely to evaluate the error of a strategy, let alone select the one with the least error.

The second key difference between the matrix mechanism and \sys is the search algorithm underlying the {\sf SELECT} step, and it is a key technical innovation of this paper. \sys uses a set of optimization routines (described in \cref{sec:optimization,sec:kron-param}) that can exploit our compact implicit workload representation.  These different optimization routines work by restricting search to different regions of the strategy space: local optimization is tractable in these regions but they still contain high quality strategies.  The output is a measurement strategy $\AA$, also represented in a compact implicit form.

Our third innovation consists of efficient techniques for measurement and reconstruction. In \matrixMech, these steps are implemented by multiplying a matrix $\A$ with the data vector $\x$ and multiplying a matrix pseudo-inverse $\A^+$ with the noisy answers $\y$, respectively. The latter inference step can be inefficient in explicit matrix form. \sys exploits the special structure of our selected measurements to speed up these steps, as described in \cref{sec:running}.

As a result of these innovations, \sys achieves high accuracy on a \emph{variety} of realistic input workloads, in both low and high dimensions.  In fact, in our experiments, we find it has higher accuracy than all prior select-measure-reconstruct techniques, even on inputs for which the prior techniques were specifically designed (e.g., it is more accurate than \HB on range queries).  We also find it is more accurate than state-of-the-art techniques outside the select-measure-reconstruct paradigm.  It achieves reasonable runtime and scales more effectively than prior work that performs non-trivial optimization (see \cref{sec:experiments} for a detailed scalability evaluation).

\paragraph*{Organization} In addition to the sections noted above, we describe our use case next, followed by background, and the end-to-end algorithm components in \cref{sec:together}, experiments in \cref{sec:experiments}, and discussion in \cref{sec:conclusion}.

\newpage
\section{Motivating use case}\label{sec:usecase}
Based on our collaboration with the U.S. Census Bureau\footnote{
{\small The Census Bureau recently announced \cite{census-url} that the test publications produced by the 2018 End-to-End Census Test would be protected by a disclosure limitation system based on differential privacy. The End-to-End Test is a prototype of the full production system to be used for the 2020 Census of Population and Housing. If the test of this disclosure limitation system is successful, then the expectation is that the publications of the 2020 Census will also be protected using differential privacy. The work discussed in this paper is part of the research and development activity for those disclosure limitation systems.}},
we use as a running example and motivating use case the differentially private release of a collection of 10 tabulations from the 2010 {\em Summary File 1 (SF1)}\cite{census-sf1}, an important data product based on the Census of Population and Housing (CPH). Statistics from SF1 are used for redistricting, demographic projections, and other policy-making.



Our workload is a subset of queries from SF1 that can be written as predicate counting queries over a \texttt{\Person} relation.  (We omit other queries involving households; for brevity we refer to our selected queries as simply SF1.)  The \texttt{\Person} relation has the following schema: six boolean attributes describing Race, two boolean attributes for Hispanic Ethnicity and Sex, Age in years between 0 and 114, and a Relationship-to-householder field that has 17 values.
Our SF1 workload has 4151 predicate counting queries, each of the form \texttt{SELECT Count(*) FROM \Person WHERE $\phi$}, where $\phi$ specifies some
combination of demographic properties (e.g. number of Persons who are Male, over 18, and Hispanic) and thus each query reports a count at the national level. These queries are on a multidimensional domain of size $2^6 \times 2 \times 2 \times 115\times 17=\num{500480}$.
The data also includes a geographic attribute encoding state (51 values including D.C.). A workload we call SF1+ consists of the national level queries in SF1 {\em as well as} the same queries at the state level for each of 51 states.
We can succinctly express the state level queries as an additional 4151 queries of the form:
\texttt{SELECT state, Count(*) FROM \Person WHERE $\phi$ GROUP BY state}. Thus, SF1+ can be represented by a total of $4151+4151=8302$ SQL queries.
%
%
The SF1+ queries are defined on a domain of size $\num{500480} \times 51=\num{25524480}$.

In addition to their SQL representation, the SF1 and SF1+ workloads can be naturally expressed in a logical form defined in \cref{sec:implicit_conjunctions}. We use \Wnt and \Wst to denote the logical forms of SF1 and SF1+ respectively.


\eat{
\section{Outline of Intro}
\begin{outline}
		\1 Problem of releasing sets of linear queries under differential privacy
		\1 has many applications
			\2 Census bureau
			\2 healthcare databases that tabulate statistics
			\2 visualizations
		\1 state of the art: either is not error optimal, or require high computational overhead and are restricted to 1- or 2-D
			\2 First class of solutions considers tables in relational form, queries in a logical form (like SQL) and adds noise calibrated to (a bound over) the sensitivity of the queries.
				\3 sensitivity of each linear query is easy to compute (1 if counting query)
				\3 but, computing sensitivity of a set of queries is NP-hard
				\3 Xiaokui, follow up, ProPER, etc.
			\2 Second class of solutions have two important differences:
				\3 consider the databases and queries in vector form expressed over the full domain of each tuple (which is exponential in the number of attributes)
				\3 rather than answering the original set of queries, these techniques (a) select a set of measurement queries, (b) measure them using Laplace mechanism and (c) reconstruct the desired answers from the measurements.
				\3 theory: [hardt-talwar] and follow ups; [Matrix Mechanism] and variants; data dependent stuff (where measurements are incomplete wrt input queries)
				\3 These are the state of the art techniques in terms of error (maybe give an example) because:
					\4 error using these strategies are an order of magnitude smaller than the naive strategy
					\4 computing sensitivity is trivial when represented in the vector form (circumventing the hardness results)
					\4 standard inference algorithms that tend to use vectors and linear algebra can readily used to reconstruct answers to input queries using noisy measurements.
				\3 But representing queries and data in vector form makes these techniques extremely inefficient, limiting their applicability
					\4 matrix mechanism algorithm too slow for standard schemas and workloads
					\4 a lot of work on specialized algorithms for special cases
			\2 other work but not workload specific
		\1 Problem:
			\2 There is no general purpose differentially private algorithm that can answer sets of statistical queries posed over a table with (a) competetive error and (b) which scales to workloads over schemas with more than a few attributes.
		\1 Contributions:
			\2 A new suite of algorithms that can answer queries with competetive error on higher dimensional tables than was possible before
			\2 scales to domains of size $10^9$
			\2 key innovation:
				\3 translate logical queries into one out of a few compact vector representation. These representations are based on outer-products or weighted marginals
				\3 For each representation, we have designed novel scalable algorithms to identify a measurement strategy for which noisy answers are derived for the database
				\3 ...
			\2 nice results
\end{outline}
}


\section{Background} \label{sec:background}

We describe below the relevant background, including the data model, logical query workloads, their corresponding vector representations and differential privacy.

\newcommand{\attset}[1]{{\cal #1}}

\subsection{Data and schema}
We assume a single-table relational schema $R(A_1 \dots A_d)$, where $attr(R)$ denotes the set of attributes of $R$. Subsets of attributes are denoted $\attset{A} \subseteq attr(R)$. Each attribute $A_i$ has a finite domain $dom(A_i)$.  The full domain of $R$ is $dom(R) = dom(A_1) \times \dots \times dom(A_d)$, containing all possible tuples conforming to $R$.  An instance $I$ of relation $R$ is a multiset whose elements are tuples in $dom(R)$. We use $N$ for $|dom(R)|$.

\vfill\null
\subsection{Logical view of queries}

Predicate counting queries are a versatile class, consisting of queries that count the number of tuples satisfying any logical predicate.
\begin{definition}[Predicate counting query] \label{def:query}
A predicate on $R$ is a boolean function $\phi: dom(R) \rightarrow  \{0,1\}$.  A predicate can be used as a counting query on instance $I$ of $R$ whose answer is
	$\phi(I)=\sum_{t\in I} \phi(t)$.
\end{definition}
A predicate corresponds to a condition in the {\tt WHERE} clause of an SQL statement, so in SQL a predicate counting query has the form: \texttt{SELECT Count(*) FROM R WHERE $\phi$}.


When a predicate $\phi$ refers {\em only} to a subset of attributes $\attset{A} \subset attr(R)$ we may annotate the predicate, writing $[\phi]_\attset{A}$.
If $[\phi_1]_\attset{A}$ and $[\phi_2]_\attset{B}$ are predicates on attribute sets $\attset{A}$ and $\attset{B}$, then their conjunction is a predicate $[\phi_1\wedge \phi_2]_{\attset{A}\cup\attset{B}}$.


We assume that each query consists of \emph{arbitrarily complex} predicates on each attribute, but require that they are combined across attributes with conjunctions. In other words, each $\phi$ is of the form $\phi = [\phi_1]_{A_1} \land \dots \land [\phi_d]_{A_d}$.  This facilitates the compact implicit representations described in \cref{sec:implicit}.  One approach to handling disjunctions (and other more complex query features) is to transform the schema by merging attributes.  We illustrate this in its application to the SF1 workload, and return to this issue in \cref{sec:conclusion}.

\begin{example}
The SF1 workload consists of conjunctive conditions over its attributes, with the exception of conditions on the six binary race attributes, which can be complex disjunctions of conjunctions (such as ``The number of Persons with two or more races'').  We simply merge the six binary race attributes and treat it like a single $2^6=64$ size attribute (called simply {\em Race}). This schema transformation does not change the overall domain size, but allows every SF1 query to be expressed as a conjunction.
\end{example}

\subsection{Logical view of query workloads} \label{sec:sub:logical_workloads}

A workload is a set of predicate counting queries. A workload may consist of queries designed to support a variety of analyses or user needs, as is the case with the SF1 workload described above.  Workloads may also be built from the sufficient statistics of models, or generated by tools that aid users in exploring data, or a combination of these analyses.  For the privacy mechanisms considered here, it is preferable for the workload to explicitly mention all queries of interest, rather than a subset of the queries that could act like a supporting view, from which the remaining queries of interest could be computed.  Enumerating all queries of interest allows error to be optimized collectively.  In addition, a workload query can be repeated, or equivalently, weighted, to express the preference for greater accuracy on that query.



\paragraph*{Structured multi-dimensional workloads}

Multi-dimens\-ional workloads are often defined in a structured form, as {\em products} and {\em unions of products}, that we will exploit later in our implicit representations.  Following the notation above, we write $\Phi=[\phi_1 \dots \phi_p]_\attset{A}$ to denote a set of $p$ predicates, each mentioning only attributes in $\attset{A}$.  For example, the following are common predicate sets defined over a single attribute $A$ of tuple $t$:\vspace{1ex} \\ \vspace{1ex}
\begin{tabular}{lll}
$I$ & $\mbox{Identity}_A$ & $= \set{t.A == a_i | a_i \in dom(A)}$  \\
$P$ & $\mbox{Prefix}_A$   & $= \{(a_1 \leq t.A \leq a_i)|a_i \in dom(A)\}$ \\
$R$ & $\mbox{AllRange}_A$ & $= \{(a_i \leq t.A \leq a_j)|a_i,a_j \in dom(A)\}$ \\
T & $\mbox{Total}_A$ & $=\set{True}$ \\
\end{tabular}

$\mbox{Identity}_A$ contains one predicate for each element of the domain.  Both $\mbox{Prefix}_A$ and $\mbox{Range}_A$ rely on an ordered $dom(A)$; they contain predicates defining a CDF (i.e. sufficient to compute the empirical cumulative distribution function), and the set of all range queries, respectively. The predicate set $\mbox{Total}_A$, consists of a single predicate, returning {\em True} for any $a \in dom(A)$, and thus counting all records.

We can construct multi-attribute workloads by taking the cross-product of predicate sets defined for single attributes, and {\em conjunctively} combining individual queries.

\begin{definition}[Product] \label{def:product-workload}
For predicate sets $\Phi=$ \\$[\phi_1 \dots \phi_p]_\attset{A}$ and $\Psi=[\psi_1 \dots \psi_r]_\attset{B}$ ($\attset{A}$ and $\attset{B}$ are disjoint), the product is a query set containing a total of $p \cdot r$ queries:
$$[\Phi \times \Psi]_{\attset{A}\cup\attset{B}} = \{ \phi_i \wedge \psi_j | \phi_i \in \Phi, \psi_j \in \Psi \}$$
\end{definition}

We describe several examples of workloads constructed from products and unions of products below.  

\begin{example}[Single query as product] \label{ex:sql_query}
A predicate counting query in the SF1 workload is: \texttt{SELECT Count(*) FROM \Person WHERE sex=M AND age < 5}.
We can express this query as a product: first, define predicate set $\Phi_1 = \set{\text{sex=}M }$ and predicate set $\Phi_2 = \set{ \text{age} < 5 }$.  The query is expressed as the product $\Phi_1 \times \Phi_2$.  (We omit Total on the other attributes for brevity.)
\end{example}

\begin{example}[\texttt{GROUP BY} query as product] \label{ex:groupby}
A \texttt{GROUP BY} query can be expressed as a product by including an Identity predicate set for each grouping attribute and a singleton predicate set for each attribute in the \texttt{WHERE} clause.
The product would also include Total for each attribute not mentioned in the query.
For example, the query \texttt{SELECT sex, age, Count(*) FROM \Person WHERE his\-panic = TRUE \\ GROUP BY sex, age} is expressed as
$\mbox{I}_{\texttt{Sex}} \times \mbox{I}_{\texttt{Age}} \times \Phi_{3}$ where $\Phi_{3} = \set{\text{hispanic=True}}$.  This product contains $2 \times 115$ counting queries, one for each possible setting of Sex and Age.
\end{example}
\eat{
\begin{example}[Marginal workloads] \label{ex:marg}
A marginal is a product defined by the product of one or more Identity predicates on selected attributes and Total on all other attributes.  For example, $\mbox{I}_{\tt Sex} \times \mbox{I}_{\tt Age} \times \mbox{I}_{\tt Hispanic}$ is a three-way marginal, which consists of $2 \times 115 \times 2$ counting queries, one for each possible setting of Sex, Age and Hispanic.  This is equivalent to a \texttt{GROUP BY} query on Sex, Age, and Hispanic with no \texttt{WHERE} clause. \gm{can we drop this example (for space)?}
\end{example}
}
\begin{example}[SF1 Tabulation as Product] \label{ex:tabulation}
Except \\ for the population total, the queries in the P12 tabulation of the Census SF1 workload \cite{census-sf1} can be described by a single product: $\mbox{I}_{\tt Sex} \times \mbox{R}_{\tt Age} $ where $ \mbox{R}_{\tt Age} $ is a particular set of range queries including $[0,114], [0,4]$, $[5,9]$, $[10,14], \dots [85,114]$.
\end{example}

\paragraph*{Unions of products}
Our workloads often combine multiple products as a union of the sets of queries in each product.  For example, the set of all three-way marginals is a union of ${ d \choose 3}$ workloads, each a product of the Identity predicate set applied to three attributes.  

The input to the algorithms that follow is a logical workload consisting of a union of products, each representing one or possibly many queries.

\begin{definition}[Logical workload] \label{def:logical_workload}
	A logical workload $\Wlog=\{q_1 \dots q_k\}$ consists of a set of products $q_i$ where each $q_i=[\Phi_{i1}]_{A_1} \times \dots \times [\Phi_{id}]_{A_d}$.
\end{definition}

\begin{example}[SF1 as union of products] \label{ex:product_terms_sf1}
The SF1 \\ workload from \cref{sec:usecase} can be represented in a logical form, denoted $\Wnt$, that consists of a union of $k=4151$ products, each representing a single query.  Because these queries are at the national level, there is a Total predicate set on the State attribute.
The logical form of the SF1+ workload, denoted \Wst, includes those products, plus an additional 4151 products that are identical except for replacing the Total on State with an Identity predicate set.  There are a total of $k=8302$ products, representing a total of $4151 + 51 \times 4151 = \num{215852}$ predicate counting queries.  While this is a direct translation from the SQL form, this representation can be reduced.  First, we can reduce to $k=4151$ products by simply adding $True$ to the Identity predicate set on State to capture the national counts.
Furthermore, through manual inspection, we found that both $\Wnt$ and $\Wst$ can be even more compactly represented as the union of 32 products---we  use $\Wnt^*$ and $\Wst^*$ to denote more compact logical forms. This results in significant space savings (\cref{ex:implicit_workload_size}) and runtime improvements.
\end{example}

\subsection{Explicit data and query vectorization} \label{sec:sub:vdata}

The vector representation of predicate counting queries (and the data they are evaluated on) is central to the select-measure-reconstruct paradigm.  
The vector representation of instance $I$ is denoted $\x_I$ (or simply $\x$ if the context is clear) and called the {\em data vector}.\footnote{When $R$ has $d$ attributes, the data vector has a multi-dimensional interpretation as a $d$-way array, or a tensor; to simplify notation we assume appropriate flattening.}  Each entry in $\x_I$ corresponds to a tuple $t\in dom(R)$ and reports the number of occurrences of $t$ in $I$.
Note that, throughout the paper, the representation of the data vector is {\em always} explicit; it is the representation of queries that will be implicit.

Every predicate counting query $\phi$ has a vector form.
\begin{definition}[Vectorized query] \label{def:vect-query}
	Given a predicate counting query $\phi$ defined on schema $R$, its vectorization is denoted $vec(\phi)$ and has an entry equal to $\phi(t)\in\{0,1\}$ for each tuple $t\in dom(R)$.
\end{definition}
The above definition immediately suggests a simple algorithm for computing $vec(\phi)$:
form a vector by evaluating $\phi$ on each element of the domain and recording the 0 or 1 output of evaluation.
(A more efficient algorithm is presented in the next section.)  Note that both the data vector and the vectorized query have size $|dom(R)|=N$.  Once a predicate query is vectorized, it can easily be evaluated by taking its dot product with the data vector: that is, $ \phi(I) = vec(\phi) \cdot \x_I $ for any instance $I$.

A single predicate counting query is represented as a vector, so a workload of predicate counting queries can be represented as a matrix in which queries are rows.  For logical workload $\Wlog$, its (explicit) matrix form is written $\W$, and the evaluation of the workload is equivalent to the matrix product $\W \x_I$.  Note that the size of the workload matrix is $m \times N$ where $m$ is the number of queries, $\x_I$ is $N \times 1$, and the vector of workload answers is $m \times 1$.
%
%

\subsection{Differential privacy}
Differential privacy is a property of a randomized algorithm that bounds the difference in output probabilities induced by changes to an individual's data. Let $\nbrs(\db)$ be the set of databases differing from $I$ in at most one record.
%
\begin{definition}[Differential Privacy~\cite{dwork2006calibrating}] \label{def:diffp}
A rand\-om\-ized algorithm $\algG$ is $(\epsilon,\delta)$-differentially private if for any instance $\db$, any $\db' \in \nbrs(\db)$, and any outputs $O \subseteq Range(\algG)$,
\[
Pr[ \algG(\db) \in O] \leq \exp(\epsilon) \times Pr[ \algG(\db') \in O] + \delta
\]
\end{definition}
We focus exclusively on $\epsilon$-differential privacy (i.e. $\delta=0$).  However our techniques also apply to a version of $\matrixMech$ satisfying approximate differential privacy ($\delta > 0)$ \cite{li2015matrix}.

The Laplace mechanism underlies the private mechanisms considered in this paper; we describe it in vector form. Let $\Lap(\sigma)^m$ denote a vector of $m$ independent samples from a Laplace distribution with mean 0 and scale $\sigma$.
\begin{definition}[Laplace mechanism, vector form] \label{prop:laplace}
Given an $m \times N$ query matrix $\A$, the randomized algorithm $\LM$ that outputs the following vector is $\epsilon$-differentially private:
$\LM(\A,\x) = \A\x + \Lap(\sigma_{\A})^m$
where $\sigma_{\A}=\frac{\Lone{\A}}{\epsilon}$.
\end{definition}
Above, $\Lone{\A}$ denotes the maximum absolute column sum norm of $\A$, shown in \cite{li2010optimizing} to be equal to the {\em sensitivity} of the query set defined by $\A$, since it measures the maximum difference in the answers to the queries in $\A$ on any two databases that differ only by a single record.

For an algorithm $\algG$ answering workload $\Wlog$, we measure error as the expected total squared error on the workload query answers, denoted $Err(\Wlog, \algG)$ .

%

%

%


\subsection{The matrix mechanism} \label{sec:sub:mm}

For a workload $\W$ in matrix form, defined on data vector $\x$, the matrix mechanism \cite{li2010optimizing} is defined in \cref{fig:alg_overview}(a). Privacy of the matrix mechanism (and thus all techniques in this paper) follows from the privacy of the Laplace Mechanism (output $\y$ of the {\sf MEASURE} step). The {\sf RECONSTRUCT} steps (inference and workload answering) perform post-processing on $\y$, so they do not degrade privacy \cite{Dwork14Algorithmic}.

We use expected total squared error as the error metric and optimization objective.  This is the same error metric proposed originally by the matrix mechanism, as well as a number of other works~\cite{li2015matrix, qardaji2013understanding, hay2010boosting, li2014data, xiao2011differential}.

\begin{definition}[Workload error under strategy] \label{def:error}
Given workload matrix $\W$ and strategy $\A$, the expected total squared error of the workload query answers is: \vspace{-1ex}
\begin{align*}
Err(\W, M\!M(\A)) &= \frac{2}{\epsilon^2}\Lone{\A}^2 || \W \A^+ ||_F^2
\end{align*}
\end{definition}


This error metric has a number of advantages: it is independent of the input data and the setting of $\epsilon$, and it can be computed in closed form \cite{li2010optimizing}.  As a result, if the workload is fixed,\footnote{E.g., the Census SF1 workload is determined for each decennial census and therefore changes only every 10 years.} the optimized strategy $\A$ can be computed once and used for multiple invocations of measure and reconstruct (i.e. on different input datasets and/or for different outputs generated with different $\epsilon$ values).

This error metric is an {\em absolute} measure of error, as opposed to a {\em relative} measure of error, which would report error normalized by the actual query answer.  The techniques in this paper are not applicable to relative error measures; the objective function of the strategy selection problem would depend on the input data, and we would need to solve for the best strategy for a workload {\em and} dataset.






\section{Implicit representations} \label{sec:implicit}


Workload matrices can be represented implicitly, in a form that is far more concise than materialized explicit workload matrices, while still allowing key operations to be performed.

\subsection{Implicitly vectorized conjunctions} \label{sec:implicit_conjunctions}

Consider a predicate defined on a single attribute, $A_1$, where $|dom(A_1)|=n_1$. This predicate, $[\phi]_{A_1}$, can be vectorized with respect to just the domain of $A_1$ (and not the full domain of all attributes) similarly to \cref{def:vect-query}.
When a predicate is formed from the conjunction of such single-attribute predicates, its vectorized form has a concise implicit representation in terms of the {\em kronecker product}, denoted $\otimes$, between vectors.  (Here we treat a length $n$ vector as an $1 \times n$ matrix.)

\begin{definition}[Kronecker product]
For two matrices $\A \in \mathbb{R}^{m \times n} $ and $\B \in \mathbb{R}^{m' \times n'}$, their Kronecker product $ \A \otimes \B \in \mathbb{R}^{m \cdot m' \times n \cdot n'} $ is:
$$ \A \otimes \B =
\begin{bmatrix}
a_{11} \B & \cdots & a_{1n} \B \\
\vdots & \ddots & \vdots \\
a_{m1} \B & \cdots & a_{mn} \B \\
\end{bmatrix}
$$
\end{definition}

\begin{theorem}[Implicit vectorization] \label{thm:implicit-vec}
Let $\phi=$ \\ $[\phi_1]_{A_1} \wedge [\phi_2]_{A_2}$ be a predicate defined by the conjunction of $\phi_1$ and $\phi_2$ on attributes $A_1$ and $A_2$.  Then, $ vec(\phi) = vec(\phi_1) \otimes vec(\phi_2)$.
\end{theorem}
While the explicit representation of $vec(\phi)$ has size $n_1 \cdot n_2$, the implicit representation, $vec(\phi_1) \otimes vec(\phi_2)$, requires storing only $vec(\phi_1)$ and $vec(\phi_2)$, which has size $n_1+n_2$.  More generally, for a predicate counting query that is a conjunction of predicates on each of the $d$ attributes, the implicit vectorized representation of the query has size just $\Sigma_{i=1}^d n_i$ while the explicit representation has size $\Pi_{i=1}^d n_i$.


\begin{example} \label{ex:imp_vec}
Recall that the workload $\Wnt$ consists of 4151 queries, each defined on a data vector of size \num{500480}.  Since explicitly vectorized queries are the same size as the domain, the size of the explicit workload matrix is $4151 \times \num{500480}$, or $8.3$GB.
Using the implicit representation, each query can be encoded using $2+2+64+115+17=200$ values, for a total of 3.3MB.
For $\Wst$, which consists of \num{215852} queries on a data vector of size $\num{25524480}$, the explicit workload matrix would require 22TB of storage. In contrast, the implicit vector representation would require 200MB.
\end{example}

\subsection{Implicitly vectorized products}

The kronecker product can naturally encode product workloads too (as in \cref{def:product-workload}).
%
%
%
Given a (logical) product $\Phi \times \Psi$, we can implicitly represent its queries as a Kronecker product of two matrices: one representing the predicates in $\Phi$ and one representing the predicates in $\Psi$.
If $\Phi$ has $p$ predicates, and the vector form of each predicate has size $n_1=|dom(A_1)|$, it is represented (explicitly) as a $p \times n_1$ matrix.  If $\Psi$ contains $r$ predicates of size $n_2=|dom(A_2)|$, it is similarly represented as an $r \times n_2$ matrix.  We can store only these matrices, implicitly representing the product as the Kronecker product:
\begin{theorem}
Given predicate sets $\Phi=[\phi_1 \dots \phi_p]_{A_1}$ and $\Psi=[\psi_1 \dots \psi_r]_{A_2}$, on attributes $A_1$ and $A_2$, the vectorized product is defined in terms of matrices $vec(\Phi)$ and $vec(\Psi)$:
$vec(\Phi \times \Psi) = vec(\Phi) \otimes vec(\Psi)$.
\end{theorem}
The size of the implicit representation is $pn_1+rn_2$, while the explicit product has size $prn_1n_2$.


\subsection{Workload encoding algorithm} \label{sec:sub:encoding}



Given as input a logical workload $\Wlog$ (as in \cref{def:logical_workload}), the $\Imp$ algorithm produces an implicitly represented workload matrix with the following form:
\begin{equation} \label{eq:wsum}
\WW_{[k]} =
\begin{bmatrix} w_1\WW_1 \\ \vdots \\ w_k\WW_k
\end{bmatrix} =
\begin{bmatrix} w_1 (\W_1^{(1)} \otimes & \dots & \otimes \W_d^{(1)})  \\ \vdots & \ddots & \vdots \\ w_k (\W_1^{(k)} \otimes & \dots & \otimes \W_d^{(k)})
\end{bmatrix}
\end{equation}
Here stacking sub-workloads is analogous to union and in formulas we will write an implicit union-of-products workload as $\WW_{[k]} = w_1\WW_1 \plus \dots \plus w_k\WW_k$.  \revision{We use blackboard bold font to distinguish an implicitly represented workload $\WW$ from an explicitly represented workload $\W$.}
%
%
%
\begin{table}[h]
\resizebox{\columnwidth}{!}{
\begin{tabular}{ll} \hline
\multicolumn{2}{l}{{\bf Algorithm 1}: $\Imp$} \\ \hline
  \multicolumn{2}{l}{ {\bf Input}: Workload $\Wlog=\{q_1 \dots q_k\}$ and weights $w_1 \dots w_k$ }\\
  \multicolumn{2}{l}{ {\bf Output}: Implicit workload $\WW_{[k]}$} \\ \hline
1.  & For each product $q_i \in \Wlog$: $q_i=[\Phi_{i1}]_{A_1} \times \dots \times [\Phi_{id}]_{A_d}$ \\
2. & \quad For each $j\in[1..d]$ \\
3. & \quad \quad compute $\W^{(i)}_j = vec(\Phi_{ij})$ \\
4. & \quad Let $\WW_i = \W_1^{(i)} \otimes  \dots \otimes \W_d^{(i)}$ \\
5. & {\bf Return:} $w_1\WW_1 \plus \dots \plus w_k\WW_k$ \\  \hline
\end{tabular} }
\end{table}
Note that line 3 of the $\Imp$ algorithm is \emph{explicit} vectorization, as in \cref{def:vect-query}, of a set of predicates on a single attribute.

\begin{example} \label{ex:implicit_workload_size}
Recall from \cref{ex:product_terms_sf1} that the \num{215852} que\-ries of \Wst can be represented as $k=8032$ products.  If $\Wst$ is represented in this factored form, the $\Imp$ algorithm returns a smaller implicit representation, reducing the 200MB (from \cref{ex:imp_vec}) to 50 MB.  If the workloads are presented in their manually factored format of $k=32$ products, the implicit representation of $\Wnt^*$ requires only 335KB, and $\Wst^*$ only 687KB.
\end{example}

%

\eat{
The SF1 specification \cite{census-sf1} does not include weights, however accuracy is a high priority for a special subset of SF1 queries; we return to this in \cref{sec:experiments}.
}



\subsection{Operations on vectorized objects} \label{sec:sub:kron-identity}
Reducing the size of the workload representation is only useful if critical computations can be performed without expanding them to their explicit representations. Standard properties of the Kronecker product~\cite{van2000ubiquitous} accelerate strategy selection and reconstruction. For strategy selection, a critical object is the matrix product $\WW^T\WW$.  When $\WW=\W_1 \otimes \dots \otimes \W_d$, then $\WW^T\WW=\W_1^T\W_1 \otimes \dots \otimes \W_d^T\W_d$.  For the inference phase of reconstruction, computing the pseudo-inverse $\AA^{+}$ is the main challenge.  If $\AA=\A_1 \otimes \dots \otimes \A_d$, then $\AA^{+}=\A_1^{+} \otimes \dots \otimes \A_d^{+}$.
%
%
Lastly, implicit strategy matrices allow for straightforward calculation of sensitivity:
\begin{theorem}[Sensitivity of Kronecker product] \label{thm:kron-sens}
Given an implicitly defined strategy matrix $\A_1 \otimes \dots \otimes \A_d$, its sensitivity is: $ || \A_1 \otimes \dots \otimes \A_d ||_1 = \prod_{i=1}^d || \A_i ||_1 $
\end{theorem}
Note that sparse matrices can also provide very compact representations but do not support key operations in implicit form; for example, $\W^T\W$ may not be sparse even if $\W$ is.

\section{Optimizing Explicit Workloads} \label{sec:optimization}

In this section we develop a solution to the strategy selection problem that works for explicitly-represented workloads and scales to modest domain sizes (of about $10^4$). This method, denoted $\optgp$, is a sub-routine used in \cref{sec:kron-param}.  

Below we state the optimization problem underlying strategy selection and introduce gradient-based optimization. We then describe a careful restriction of the strategy space required to make gradient-based optimization scalable and effective, leading to the definition of $\optgp$. 

\subsection{The optimization problem} \label{sec:sub:problem}

Our goal is to find a strategy $\A$ that (a) \emph{supports} the input workload $\W$ while (b) offering minimum expected total squared error as per Definition~\ref{def:error}.  Strategy $\A$ supports a workload $\W$ if and only if every query in $\W$ can be expressed as a linear combination of the queries in $\A$, which occurs whenever $ \W = \W \A^+ \A $ ~\cite{li2015matrix}. The expected total squared error of the workload using a strategy with sensitivity 1\footnote{Li et al \cite{li2010optimizing} showed that error-optimal strategy matrices have equal $L_1$ column norm, which can be normalized to 1.} is equal to $ || \W \A^+ ||_F^2 $, where $||\cdot||_F$ is the Frobenius norm. The resulting constrained optimization problem is: 

\begin{problem}\label{prob:direct}
Given an $m\times n$ workload matrix $ \W $:
\begin{equation} \label{eq:opt}
\begin{aligned}
& \underset{\A \in \mathbb{R}^{p \times n}}{\text{minimize}}
& & || \W \A^+ ||_F^2 \\
& \text{subject to} & & \W \A^+ \A = \W \mbox{ , } || \A ||_1 \leq 1
\end{aligned}
\end{equation}
\end{problem}
This optimization problem is difficult to solve exactly. It has many variables and is not convex, both the objective function and constraints involve $\A^+$, which can be slow to compute, and, in addition, the constraint on $ || \A ||_1 $ is not differentiable.  Finally, $ || \W \A^+ ||_F^2 $ has points of discontinuity near the boundary of the constraint $ \W\!\A^+\!\A = \W $. This problem was originally formulated as a rank-constrained semi-definite program \cite{li2010optimizing}, which will converge to the global optimum, but requires $O(m^4(m^4+N^4))$ time, making it infeasible for practical cases. 

Gradient-based numerical optimization techniques can be used to find locally optimal solutions.  These techniques begin by guessing a solution $\A_0$ and then iteratively improving it using the gradient of the objective function to guide the search. The process ends after a number of iterations are performed, controlled by a stopping condition based on improvement of the objective function.  A direct application of gradient-based optimization methods does not work due to the constraints, and so a projected gradient method must be used instead.   Even without the constraints, gradient-based optimization is slow, as the cost of computing the objective function for general $\A$ is $O(N^3)$, e.g. requiring more than $6$ minutes for $N=8192$.

\eat{
\subsection{Gradient-based optimization} \label{sec:sub:gradient}
\am{This should not be a subsection of its own, but rather a continuation of the previous discussion. Previous para: Exact techniques do not work. The next paras: Gradient based numerical optimization also does not work. Therefore, we need $\optgp$. Also, given that $\optgp$ reduces the cost from $n^3$ to only $n^2p$, this part should make a bigger deal about the fact that constraints are not satisfied (which impact error and privacy) rather than the runtime being prohibitive. You can move the runtime analysis to the section 4.3.} \gm{OK. Note that at the end of section we explain that the performance improvement is larger than the $n/p$ asymptotic factor suggests.}
Our method adapts gradient-based numerical optimization techniques to find locally optimal solutions.  Gradient-based algorithms begin by guessing a solution $\A_0$ and then iteratively improving it using the gradient of the objective function to guide the search.  As is the case with many non-convex optimization problems, the initialization can impact the quality of the solution.  We expect to run a number of restarts of the optimization, each beginning from a random initialization, in order to avoid unfavorable local minima.  For each initialization, a number of iterations are performed, controlled by a stopping condition based on improvement of the objective function.  

The runtime of gradient-based optimization is: 
$\mbox{\sc \#restarts}*\mbox{\sc \#iter}*(\mbox{\sc cost}_{grad}+\mbox{\sc cost}_{obj})$,
where $\mbox{\sc cost}_{grad}$ is the cost of computing the gradient, $\mbox{\sc cost}_{obj}$ is the cost of computing the objective function, {\sc \#restarts} is the number of random restarts, and {\sc \#iter} is the number of iterations.  The efficiency of this class of techniques is largely determined by $\mbox{\sc cost}_{grad}$ and $\mbox{\sc cost}_{obj}$ so we report the asymptotic complexity of these. End-to-end performance will be assessed empirically in Section \ref{sec:experiments}.  

For strategy $\A$, the objection function, denoted $C(\A)$, and the gradient function, denoted $ \frac{\partial C}{\partial \A} $, are defined as: 
\begin{equation} \label{eq:obj}
C(\A) = || \W \A^+ ||_F^2 = tr[(\A^T \A)^+ (\W^T \W)]
\end{equation}
\begin{equation} \label{eq:grad}
\begin{aligned}
\frac{\partial C}{\partial \A} &= -2 \A (\A^T \A)^+ (\W^T \W) (\A^T \A)^+
\end{aligned}
\end{equation}
%
Assuming $ \W^T \W $ has been precomputed, the complexity of evaluating $C(\A)$ doesn't depend on the number of queries in $\W$ since $ \W^T \W $ is always $ n \times n$.  \ryref{$\W^T \W$ only needs to be computed once, whereas the objective and gradient need to be evaluated many times.  For large highly structured workloads such as all range queries, $\W^T \W$ is much smaller than $\W$, and it can be computed directly without $\W$.  To support these types of workloads which could be impossible to represent in matrix form, we allow $\W^T \W$ to be passed directly to $\optgp$ in place of $\W$.}{not exactly sure where this stuff belongs}


Unfortunately, computing the gradient and objective is prohibitively  expensive for Problem \ref{prob:direct}. It is $O(N^3)$, and in practice computing the objective takes about 45 seconds (for $N=4096$) and more than 6 minutes (for $N=8192$).  It is typically impossible to use gradient-based optimization with such a costly objective function, and in addition, we have not considered the constraints in Problem \ref{prob:direct}. We address both of these obstacles next.

}

\subsection{Parameterized optimization: $\optgp$} \label{sec:sub:generalpurpose}
We now present an algorithm to solve Problem~\ref{prob:direct} by judiciously restricting the search space of the optimization problem. Recall that our goal is to search over $\A$ that support the workload $\W$. The key observation we make is the following: if $\A$ contains one query that counts the number of records in the database for each domain element in $dom(R)$ -- i.e. $\A$ contains a scaled identity matrix -- then any workload $\W$ is supported by $\A$. Of course $\A = \I$ is not a good strategy for many workloads (e.g., this has poor error for a workload encoding prefix queries). Hence, we search over strategies that contain a scaled identity matrix in addition to $p$ extra rows, as defined below.



\begin{definition}[$p$-Identity strategies] \label{def:gp} 
Given a $p \times N$ matrix of non-negative values $\matr{\Theta}$, the $p$-Identity strategy matrix $\A(\matr{\Theta})$ is defined as follows: \vspace{-3ex}
\begin{equation*} \label{eq:reparam}
A(\matr{\Theta}) = \begin{bmatrix} \I \\ \matr{\Theta} \end{bmatrix} \D
\end{equation*}
where $\I$ is the identity matrix and $ \D = diag(\vect{1}_N + \vect{1}_p \matr{\Theta})^{-1} $.
\end{definition}
Above, $\D$ is a diagonal matrix that scales the columns of $\A(\matr{\Theta})$ so that $ || \A ||_1=1$.
The $p \times N$ values in $\matr{\Theta}$ determine the weights of the $p$ queries as well as the weights on the identity queries (where a lower weight query will be answered with greater noise).  

\begin{example}
For $p=2$ and $N=3$, we illustrate below how $\A(\matr{\Theta})$ is related to its parameter matrix, $\matr{\Theta}$. 

\vspace{-3ex}
{\small
\begin{align*} 
\matr{\Theta} = \begin{bmatrix} 1 & 2 & 3 \\ 1 & 1 & 1 \end{bmatrix} & & &
\A(\matr{\Theta}) = \begin{bmatrix} 0.33 & 0 & 0 \\ 0 & 0.25 & 0 \\ 0 & 0 & 0.2 \\ 0.33 & 0.5 & 0.6 \\ 0.33 & 0.25 & 0.2 \end{bmatrix}
\end{align*} }
\end{example} 

For this class of parameterized strategies, the resulting optimization problem is stated below; we use $\optgp$ to denote the operator that solves this problem.

\begin{problem}[parameterized optimization]\label{prob:gp} 
Given \\ workload matrix $ \W $ and hyper-parameter $p$: \vspace{-1ex}
\begin{equation*} 
\begin{aligned}
& \underset{\{\A=\A(\matr{\Theta}) \:|\: \matr{\Theta} \in \mathbb{R}_+^{p \times n}\}}{\text{minimize}}
& & || \W \A^+ ||_F^2 \\
\end{aligned}
\end{equation*}
\end{problem}

%

This parameterization was carefully designed to provide the following beneficial properties: 

\vspace{1ex}
\noindent {\bf \em Constraint resolution} \quad Problem \ref{prob:gp} is a simpler optimization problem than Problem \ref{prob:direct} because it is unconstrained: $ \W \!\A^+\! \A = \W $ and $ || \A ||_1 \leq 1 $ are satisfied for all $\A=\A(\matr{\Theta})$.

\vspace{1ex}
\noindent {\bf \em Expressiveness} \quad Considering only $p$-Identity strategy matrices could mean that we omit from consideration the optimal strategy, but it also reduces the number of variables, allowing more effective gradient-based optimization.  The expressiveness depends on the parameter $p$ used to define matrices $\A(\matr{\Theta})$, which is an important hyper-parameter.  
In practice we have found $p \approx \frac{n}{16}$ to provide a good balance between efficiency and expressiveness for complex workloads (such as the set of all range queries). For less complex workloads, an even smaller $p$ may be used.  



\vspace{1ex}
\noindent {\bf \em Efficient gradient, objective, and inference} \quad 
%
To a first approximation, the runtime of gradient-based optimization is
$\mbox{\sc \#restarts}*\mbox{\sc \#iter}*(\mbox{\sc cost}_{grad}+\mbox{\sc cost}_{obj})$,
where $\mbox{\sc cost}_{grad}$ is the cost of computing the gradient, $\mbox{\sc cost}_{obj}$ is the cost of computing the objective function, {\sc \#restarts} is the number of random restarts, 
and {\sc \#iter} is the number of iterations per restart.  
For strategy $\A$, the objection function, denoted $C(\A)$, and the gradient function, denoted $ \frac{\partial C}{\partial \A} $, are defined: 
\begin{align}
C(\A) &= || \W \A^+ ||_F^2 = tr[(\A^T \A)^+ (\W^T \W)] \label{eq:obj} \\  
\frac{\partial C}{\partial \A} &= -2 \A (\A^T \A)^+ (\W^T \W) (\A^T \A)^+ \label{eq:grad}
\end{align}

%
Note that the above expressions depend on $\W$ through the matrix product $\W^T\W$, which can be computed once and cached for all iterations and restarts; we therefore omit this cost from complexity expressions.  It always has size $N \times N$ regardless of the number of queries in $\W$.  For highly structured workloads (e.g all range queries), $\W^T\W$ can be computed directly without materializing $\W$, so we allow $\optgp$ to take $\W^T\W$ as input in these special cases. 

For general $\A$, the runtime complexity of computing $C(\A)$ and $ \frac{\partial C}{\partial \A} $ is $O(N^3)$. By exploiting the special structure of $\A(\matr{\Theta})$, we can reduce these costs to $ O(pN^2) $:


\begin{restatable}[Complexity of $\optgp$]{theorem}{thmoptgp}\label{thm:optgp}
Given any $p$-Id\-ent\-ity strategy $\A(\matr{\Theta})$, both the objective function $C(\A(\matr{\Theta}))$
and the gradient $ \frac{\partial C}{\partial \A}$ can be evaluated in $ O(pN^2) $ time.  	
\end{restatable}

The speedup resulting from this parameterization is often much greater in practice than the $\frac{N}{p}$ improvement implied by the theorem.  When $N = 8192$, computing the objective for general $\A$ takes $>6$ minutes, while it takes only $1.5$ seconds for a $p$-Identity strategy: a $240\times$ improvement.  
Nevertheless, $\optgp$ is only practical for modest domain sizes ($N \sim 10^4$). For multi-dimensional data we do not use it directly, but as a subroutine in the algorithms to follow.


\section{Optimizing Implicit Workloads} \label{sec:kron-param}

\begin{table*}[h]
\caption{\label{tbl:opt-summary} Summary of optimization operators: input and output types, and the complexity of objective/gradient functions.}
\resizebox{\textwidth}{!}{
\begin{tabular}{ll|c|rcl|c|l}
\multicolumn{2}{c}{\bf Definition} & 
\multicolumn{1}{c}{{\bf Output strategy}} & 
\multicolumn{3}{c}{\bf Optimization Operator} & 
\multicolumn{1}{c}{{\bf Input workload}} & 
{\bf Complexity} \\  \hline\hline
\textsection \ref{sec:sub:generalpurpose} & Problem \ref{prob:gp} & 
$p$-Identity matrix & 
$\A(\matr{\Theta})$ & $\leftarrow$ & $\optgp(\W,p)$ & 
Any explicit	 $\W\!$	& 
$N^2 p$ \\ \hline

\textsection \ref{sec:sub:opt1} & Definition \ref{def:opt1}  & 
Product, $A(\matr{\Theta})$ terms &
$\AA$ & $\leftarrow$ & $\optk(\WW, \vec{p})$ 	& 
Single product	& 
$\sum_{i=1}^d n_i^2 p_i$ \\ \hline

\textsection \ref{sec:sub:unionkron} & Problem \ref{prob:sum-kron}  & 
Product, $A(\matr{\Theta})$ terms &
$\AA$ &  $\leftarrow$ &  $\optk(w_1\WW_1 + \dots + w_k\WW_k,\vec{p})$ & 
Union of products & 
$k \sum_{i=1}^d n_i^2 p_i$ \\ \hline

\textsection \ref{sec:sub:unionkron} & Definition ~\ref{def:optp}  & 
Union of products & 
$\AA_1 + \dots + \AA_l$ & $\leftarrow$ & $\optkk(\WW_{[k_1]}, \dots, \WW_{[k_l]}, \vec{p})$	&  
Union of products &  
$k \sum_{i=1}^d n_i^2 p_i $ \\ \hline
 
\textsection \ref{sec:sub:marginals} & Problem ~\ref{prob:marginal}  & 
Weighted marginals & 
$\mathbb{M}(\vect{\theta})$ &  $\leftarrow$  & $\optm(w_1\WW_1 + \dots + w_k\WW_k)$ & 
Union of products & 
$4^d$ \\ \hline \hline
\end{tabular} }
\end{table*}

We present next a series of optimization techniques for multi-dimensional workloads, exploiting the implicit workload representations presented in Section \ref{sec:implicit}.  One of the main ideas is to decompose a strategy optimization problem on a multi-dimensional workload into a sequence of optimization problems on individual attributes.  The optimization operators $\optk$, $\optkk$, and $\optm$ are each variants described below and are summarized in \cref{tbl:opt-summary}. 

\subsection{Optimizing product workloads} \label{sec:sub:opt1}

Recall that $\WW = \W_1 \otimes \dots \otimes \W_d $, for an implicit workload that is the product of $d$ sub-workloads. If $ \W_i $ is defined with respect to a domain of size $n_i$, then $ \WW$ is defined on a total domain of size of $N = \prod_{i=1}^d n_i $.  We perform strategy optimization directly on this implicit product representation by decomposing the optimization problem into a series of explicit optimizations on the sub-workloads.
\begin{definition}[$\optk$] \label{def:opt1}
Given workload $\WW$ and parameter vector $\vec{p}=\langle p_1 \dots p_d \rangle$, the operator $\optk(\WW,\vec{p})$ applies $\optgp$ to each sub-workload and returns a product strategy:
\begin{align*}
\optk(\WW,\vec{p}) & = \optk(\W_1 \otimes \dots \otimes \W_d, \vec{p})   \\
& \buildrel \text{d{}ef}\over = 
			 \optgp(\W_1,p_1) \otimes \dots \otimes \optgp(\W_d,p_k) \\
			& = \A_1 \otimes \dots \otimes \A_d  =  \AA
\end{align*}
\end{definition}
Therefore $\optk$ merely requires solving $d$ independent instances of $\optgp$ and the resulting output strategy is the product of $d$ distinct $p_i$-Identity strategies.  This decomposition has a well-founded theoretical justification. Namely, if we restrict the solution space to a (single) product strategy, so that $\A$ has the form $\AA = \A_1 \otimes \dots \otimes \A_d$, then the error of the workload under $\AA$ decomposes into the product of the errors of its sub-workloads under corresponding sub-strategies. Thus overall error is minimized when $ Err(\W_i,\A_i) $ is minimized for each $i$ and it follows that the $\optk$ program minimizes the correct objective function.
\begin{restatable}[Error decomposition]{theorem}{thmerrdecomp}\label{thm:err-decomp}
Given a workload $\WW = \W_1 \otimes \dots \otimes \W_d $ and a sensitivity $1$ strategy $\AA = \A_1 \otimes \dots \otimes \A_d$, the error is proportional to: \vspace{-1ex}
$$ || \WW \AA^+ ||_F^2 = \prod_{i=1}^d || \W_i \A_i^+ ||_F^2 $$
\end{restatable}
The cost of each iteration in $\optk(\WW)$ is the sum of costs of $d$ independent optimizations of the sub-workloads $\W_i$.  Since the cost of each iteration in $\optgp$ is $O(p_i n_i^2)$, the cost for $\optk(\WW)$ is $O(\sum p_i n_i^2)$.  Note that if $\WW$ were represented explicitly, each iteration in $\optgp$ would take $O(\prod_ip_i n_i^2)$.

%

\subsection{Optimizing unions of product workloads} \label{sec:sub:unionkron}

We now define three approaches for optimizing implicit workloads that are (weighted) unions of products.  Each approach restricts the strategy to a different region of the full strategy space for which optimization is tractable (see \cref{tbl:opt-summary}).  The first computes a strategy consisting of a single product; it generalizes $\optk$.  The second, $\optkk$, can generate strategies consisting of unions of products.  The third, $\optm$, generates a strategy of weighted marginals.  

\vspace{1ex}
\noindent {\bf \em Single-product output strategy} \quad 
For weighted union of product workloads, if we restrict the optimization problem to a single product strategy, then the objective function decomposes as follows:
\begin{theorem}
Given workload $\WW_{[k]}=w_1\WW_1 \plus \dots \plus w_k\WW_k$ and strategy matrix $\AA = \A_1 \otimes \dots \otimes \A_d$, workload error is: \vspace{-1ex}
\begin{equation} \label{eq:unionkron}
|| \WW_{[k]} \AA^+ ||_F^2 = \sum_{j=1}^k w_j^2 \prod_{i=1}^d || \W^{(j)}_i \A_i^+ ||_F^2 
\end{equation}
\end{theorem} 
This leads to the following optimization problem:

\begin{problem}[Union of product optimization]\label{prob:sum-kron}
For a workload $\WW_{[k]}=w_1\WW_1 \plus \dots \plus w_k\WW_k$ and parameter vector $\vec{p}=\langle p_1 \dots p_d\rangle$, let: \vspace{-2ex}
\begin{equation*} 
\begin{aligned}
\A_i &= \underset{\{\A_i(\matr{\Theta}_i) \:|\: \matr{\Theta}_i \in \mathbb{R}_+^{p_i \times n}\}}{\text{minimize}}
& & \sum_{j=1}^k w_j^2 \prod_{i'=1}^d || \W^{(j)}_{i'} \A_{i'}^+ ||_F^2 
\end{aligned}
\end{equation*}
and form final solution strategy as $\AA = \A_1 \otimes \dots \otimes \A_d$.
\end{problem}
\noindent When $k=1$, Problem \ref{prob:sum-kron} returns the same solution as \cref{def:opt1}, so we use matching notation and allow the $\optk$ operator to accept a single product or a union of products, as shown in \cref{tbl:opt-summary}.  Problem~\ref{prob:sum-kron} is a coupled optimization problem, and we use a block method that cyclically optimizes $\A_1, \dots, \A_d$ until convergence.  We begin with random initializations for all $\A_i$. We optimize one $\A_i$ at a time, fixing the other $\A_{i'} \neq \A_i$ using $\optgp$ on a carefully constructed surrogate workload $\hat{\W}_i$ (equation~\ref{eq:surrogate}) that has the property that the error of any strategy $\A_i$ on $\hat{\W}_i$ is the same as the error of $\AA$ on $\WW$.  Hence, the correct objective function is optimized. 
{\small \begin{equation} \label{eq:surrogate}
\begin{aligned} 
\hat{\W}_i = \begin{bmatrix} c_1 \W_i^{(1)} \\ \vdots \\ c_k \W_i^{(k)} \end{bmatrix}
& &
c_j = w_j \prod_{i' \neq i} || \W_{i'}^{(j)} \A_{i'}^+ ||_F
\end{aligned}
\end{equation}}
The cost of running this optimization procedure is determined by the cost of computing $ \hat{\W}_i^T \hat{\W}_i $ and the cost of optimizing it, which takes $ O(n_i^2 (p_i+k)) $ and $ O(n_i^2 p_i \cdot \text{\sc \#iter}) $ time respectively (assuming each $(\W^T \W)_i^{(j)}$ has been precomputed). As before, this method scales to arbitrarily large domains as long as the domain size of the sub-problems allows $\optgp$ to be efficient.  



\vspace{1ex}
\noindent {\bf \em Union-of-products output strategy} \label{sec:sub:unionkron2} \quad 
For certain workloads, restricting to solutions consisting of a single product, as $\optk$ does, excludes good strategies.  This can happen for a workload like 
$W = (R \times T) \cup (T \times R)$,
for which choosing a single product tends to force a suboptimal pairing of queries across attributes. Unfortunately, we cannot optimize directly over union-of-product strategies because computing the expected error is intractable.  Nevertheless, we can use our existing optimization methods to generate high-quality union-of-product strategies. This operator takes as input a weighted union of products, partitioned into $l$ subsets. It optimizes each individually using $\optk$ and combines the resulting output strategies to form a strategy consisting of a union of $l$ products.  Below we use $K=k_1 + \dots + k_l$ (and recall notation $\WW_{[k]}$ from \cref{sec:sub:encoding}):
\begin{definition}[$\optkk$] \label{def:optp} Given a workload, $ \WW_{[K]} =$ \\ $\WW_{[k_1]} + \dots + \WW_{[k_l]} $, and parameter vector $\vec{p}$, the optimization routine $\optkk$ returns the union of strategies defined below: 
\begin{align*}
\optkk(\WW_{[K]},\vec{p}) & 
\buildrel \text{d{}ef}\over = 
\optk(\WW_{[k_1]},\vec{p})  + \dots + \optk(\WW_{[k_l]},\vec{p}) \\
& = \AA_1 + \dots + \AA_l
\end{align*}
\end{definition}

This definition could easily be extended so that each $ \AA_i $ gets a different fraction of the privacy budget, and so that each call to $\optk$ gets a different parameter vector.

\subsection{Optimized marginal strategies} \label{sec:sub:marginals}

Although $\optk$ or $\optkk$ can be used to optimize workloads consisting of marginals, we now describe a third optimization operator, $\optm$, which is especially effective for marginal workloads (but is applicable to any union of product workload).  A single marginal can be specified by a subset of attributes $S \subseteq [d]$ and can be expressed as the product $\A_1 \otimes \dots \otimes \A_d $ where $\A_i = \I $ if $ i \in S $ and $ \A_i = \T $ otherwise.  
Since there are $2^d$ subsets of $[d]$, a set of weighted marginals can be characterized by a vector $ \vect{\theta}$ of $2^d$ non-negative weights where $\theta_1$ is the weight on the 0-way marginal (i.e., the total query) and $\theta_{2^d}$ is the weight on the $d$-way marginal (i.e., the queries defining the full contingency table). We use $\mathbb{M}(\vect{\theta})$ to denote the matrix which stacks each of these $2^d$ weighted marginals.  Each marginal has sensitivity $1$ so the total sensitivity of $\mathbb{M}(\vect{\theta})$ is $ \sum \theta_i $.  We resolve the sensitivity constraint from Problem \ref{prob:direct} by moving it into the objective function, and we resolve the other constraint by forcing $ \theta_{2^d} $ to be strictly positive.  The resulting optimization problem is given below: 
%
%
\begin{problem}[Marginals optimization]\label{prob:marginal}
Given a \\ workload, $\WW_{[k]} = $ $w_1\WW_1$ $ \plus \dots \plus w_k\WW_k$, let:
\vspace{-2ex}
\begin{equation*} 
\begin{aligned}
\vect{\theta} =  \underset{\vect{\theta} \in \mathbb{R}_+^{2^d}; \theta_{2^d} > 0}{\text{minimize}}
& & \Big( \sum \theta_i \Big)^2 || \WW_{[k]} \mathbb{M}(\vect{\theta})^+ ||_F^2
\end{aligned}
\end{equation*}
\end{problem}
Without materializing $\mathbb{M}(\vect{\theta})$ we can evaluate the objective function and its gradient by exploiting the structure of $\mathbb{M}(\vect{\theta})$.  For strategies of the form $\mathbb{M}(\vect{\theta})$, $ (\mathbb{M}^T \mathbb{M})^+ $ can be written as the weighted sum of $ 2^d $ kronecker products, where each sub-matrix is either $ \I $ or $\matr{1} = \T^T \T $, and we can efficiently find the $2^d$ weights that characterize the inverse by solving a (sparse) linear system of equations.  The objective function only depends on $\WW_{[k]}$ through the trace and sum of $ (\W^T \W)^{(j)}_i $.  Thus, these statistics can be precomputed for the workload and the objective function can be evaluated very efficiently. The cost of this precomputation is linear in $k$, but subsequent to that, evaluating the objective and gradient only depends on $d$, and not $n_i$ or $k$.  Specifically, the time complexity of evaluating the objective and gradient is $O(4^d)$ -- quadratic in $2^d$, the size of $\vect{\theta}$.  

\section{The \sys Algorithm} \label{sec:together}

The complete \sys algorithm is described in \cref{fig:alg_overview}(b). Here we explain $\metaopt$, efficient {\sf MEASURE} and {\sf RECONSTRUCT} methods, and conclude with a privacy statement.

\subsection{The $\metaopt$ strategy selection algorithm} \label{sec:sub:prog_space}

Using the optimization operators defined in previous sections, we now define $\metaopt$, our fully-automated strategy selection algorithm.  Predicting which optimization operator will yield the lowest error strategy requires domain expertise and may be challenging for complex workloads.  Since our operators are usually efficient, we simply run multiple operators, keeping the output strategy that offers least error.  We emphasize that strategy selection is independent of the input data and does not consume the privacy budget.

The algorithm below takes as input: implicit workload $\WW$, operator set $\mathcal{P}$, and the maximum number of restarts, $S$. \vspace{-2ex} \\
\begin{table}[h]
\resizebox{\columnwidth}{!}{
\begin{tabular}{ll} \hline
\multicolumn{2}{l}{{\bf Algorithm 2}: $\metaopt$} \\ \hline
  \multicolumn{2}{l}{ {\bf Input}: implicit workload $\WW$, operator set $\mathcal{P}$, max-restarts $S$} \\
  \multicolumn{2}{l}{ {\bf Output}: implicit strategy $\AA$} \\ \hline

1.  & $best$ = $(\I, error_{\I})$ \\
2.   & {\bf For} random starts $[1..S]$: \\
3.  & \quad {\bf For} each operator $P_i \in \mathcal{P}$:  \\
4.  &  \qquad $(\AA_i, error_i)=P_i(\WW)$\\
5. & \qquad {\bf if} $error_i < best$ {\bf emit} $\AA_i$ and {\bf update} {\em best}. \\
\hline
\end{tabular} }
\end{table}

We instantiate $\metaopt$ with an operator set consisting of: $\optk(\WW,\vec{p})$, $\optkk(g(\WW),\vec{p})$, and $\optm(\WW)$.  ($\optgp$ does not appear here explicitly because it is called by $\optk$ and $\optkk$).  We use the following convention for setting the $p$ parameters: if an attribute's predicate set is contained in $T \cup I$, we set $p=1$ (this is a fairly common case where more expressive strategies do not help), otherwise we set $p=n_i/16$ for each attribute $A_i$ with size $n_i$.  In $\optkk$ above, $g$ forms two groups from the unioned terms in $\WW$.

Extensions to the above algorithm---e.g., cost-based exploration of the operator space, or an extended operator set with alternative parameter settings---could improve upon the already significant improvements in accuracy we report in our experimental evaluation and are left for future work.


\subsection{Efficient {\sf MEASURE} and {\sf RECONSTRUCT}} \label{sec:running}

The implicit form of the output strategies enables efficiency for the \textbf{measure} and \textbf{reconstruct} stages.
Recall from~\cref{sec:kron-param} that $\optk$ returns a product strategy and $\optkk$ returns a union of products, and in both cases, the $d$ terms in each product are $p$-Identity matrices. Similarly, $\optm$ returns a union of products consisting of the marginals building blocks $\T$ and $\I$.

To exploit this structure to accelerate the {\sf MEASURE} phase we define an efficient $\mult(\AA, \x)$ operation (as in \cref{fig:alg_overview}(b)) for $\AA = \A_1 \otimes \dots \otimes \A_d$ (although it is easily extended to a union of kronecker products.)  The key property of Kronecker products that we need is given in Equation~\ref{eq:implicitkron}:
\begin{equation} \label{eq:implicitkron}
(\B \otimes \C) \operatorname{flat}(\X) = \operatorname{flat}(\B \X \C^T)
\end{equation}
where $\operatorname{flat}(\cdot)$ ``flattens'' a matrix into a vector by stacking the (transposed) rows into a column vector. The expression on the right is computationally tractable, as it avoids materializing the potentially large matrix $\B \otimes \C$.  Matrices $\B$ and $\C$ need not be explicitly represented if we can compute matrix-vector products with them.  Thus, by setting $ \B \leftarrow \A_1 \otimes \dots \otimes \A_{d-1} $ and $\C \leftarrow \A_d$ we can compute matrix-vector products for matrices of the form $ \A_1 \otimes \dots \otimes \A_d $ efficiently. When $d>2$, $\B$ would never be materialized; instead we repeatedly apply Equation~\ref{eq:implicitkron} to get a representation of $\B$ that can be used to compute matrix-vector products.

Assuming for simplicity that $\A_i \in \mathbb{R}^{n \times n}$ for all $i$, the space and time complexity of computing $\mult(\AA, \x)$ using this procedure is $O(n^d)$ and $O(d n^{d+1})$ respectively, where $n^d$ is the size of the data vector.  Using an explicit matrix representation would require $O(n^{2d})$ time and space.

Related techniques allow {\sf RECONSTRUCT} to be accelerated because $\invert(\AA, \y)$ can be defined as $ \mult(\AA^+, \y) $ where $\AA^+$ is the pseudo inverse of $\AA$. For strategies produced by $\optk$ and $\optm$, the pseudo inverse can be computed efficiently in implicit form.
For $\optk$, we use the identity in \cref{sec:sub:kron-identity}: $ (\A_1 \otimes \dots \otimes \A_d)^+ = \A_1^+ \otimes \dots \otimes \A_d^+ $.  For $\optm$, the pseudo inverse of the strategy $ \MM(\vect{\theta}) $ is $ \MM^+ = (\MM^T \MM)^+ \MM^T $, where $\MM^T$ is a (transposed) union of Kronecker products, and $(\MM^T \MM)^+$ is a sum of Kronecker products.  Thus, we can define an efficient routine for $\invert$ for these strategies.
Unfortunately, we are not aware of an efficient method for computing the pseudo inverse of a strategy produced by $\optkk$, but we can still perform inference by using an iterative algorithm like LSMR~\cite{fong2011lsmr} to solve the least squares problem. This algorithm only requires as input a routine for computing matrix-vector products on $\AA$ and $ \AA^T$, which we can do efficiently using $ \mult $.

\subsection{Privacy statement}
We conclude with a statement of the privacy of \sys. The $\Imp$ and $\metaopt$ steps of \sys do not use the input $\x$.  The Laplace Mechanism is used to compute $\xhat$ from $\x$ and the privacy of \sys follows from privacy properties of the Laplace Mechanism \cite{dwork2006calibrating} and the well-known post-processing theorem \cite{Dwork14Algorithmic}.
\begin{theorem}[Privacy]
	The \sys algorithm is $\epsilon$-differentially private.
\end{theorem}

\revision{
\section{Experiments} \label{sec:experiments}

In this section we evaluate the accuracy and scalability of \sys.  We begin with details on the experimental setup. In \cref{sec:sub:accuracy} we perform a comprehensive comparison of \sys with competing algorithms, showing that it consistently offers lower error and works in a significantly broader range of scenarios than other algorithms.  In \cref{sec:sub:scalability}, we study the scalability of HDMM compared with the subset of methods that perform search over a general strategy space, showing the HDMM scales more favorably than competitors and can efficiently support high-dimensional cases.  
\newcommand{\resultrow}[9]{#1 & \NS & #6 & \emph{\textbf{1.0}} & #2 & #3 & #5 & #4 & #9 & #7 & #8 \\}

\begin{table*}[t!]
\caption{
\label{tab:main} Error ratios of various algorithms on low and high dimension datasets/workloads with $\epsilon=1.0$. Algorithms labeled \NA\:are not applicable for the given configuration; algorithms labeled \NS\:are not scalable to the given configuration.}
\centering
\resizebox{\textwidth}{!}{
\begin{tabular}{|c|c|c||ccccc|ccccc|cc|}
\hline
\multicolumn{3}{|c||}{\textbf{Configuration}} & \multicolumn{5}{c|}{\textbf{General-Purpose Algorithms}} & \multicolumn{5}{c|}{\textbf{Low-D Range Query Algorithms}} & \multicolumn{2}{c|}{\textbf{High-D Algorithms}} \\
 \multicolumn{1}{|c}{\textbf{Dataset}}                 & \multicolumn{1}{c}{\textbf{Domain / Dimensions}}                            & \textbf{Workload}            &
$\textbf{\Identity}$ & $\textbf{\LMW}$ & \textbf{\matrixMech} & \textbf{\LRM} & \textbf{\sys} &
\textbf{\Privelet} & \textbf{\HB} & \textbf{\Quadtree} & \textbf{\GreedyH} & \textbf{\DAWA} &
\textbf{\DataCube} & \textbf{\PrivBayes} \\ \hline
\multirow{3}{*}{\textbf{\Patent}} & \multirow{3}{*}{$\mathbf{1024}$} & \textbf{\FixedWidthRange} &
\resultrow{\textbf{1.25} & 7.06}{2.59}{1.48}{\textbf{1.25}}{\NA}{3.21}{\NA}{\NA}{2.45}
& & \textbf{\PrefixOneD} &
\resultrow{3.34 & 151}{1.80}{\textbf{1.34}}{1.49}{\NA}{2.44}{\NA}{\NA}{2.96}
& & \textbf{\PermutedRange}&
\resultrow{2.36 & 877000}{10.57}{3.35}{\textbf{2.16}}{\NA}{\NS}{\NA}{\NA}{\NS} \cline{1-3}
\multirow{2}{*}{\textbf{\Taxi}} & \multirow{2}{*}{$\mathbf{256 \times 256}$} & \textbf{\PrefixIdentity} &
\resultrow{\textbf{1.44} & 65.0}{6.11}{4.05}{\NS}{4.71}{\NS}{\NA}{\NA}{\NS}
& & \textbf{\PrefixTwoD} &
\resultrow{4.75 & 2422}{3.14}{2.03}{\NS}{\textbf{1.95}}{\NS}{\NA}{\NA}{\NS} \hline
\multirow{2}{*}{\textbf{\Census}} & \multirow{2}{*}{$\mathbf{2 \times 2 \times 64 \times 17 \times 115 \times 51}$} & \textbf{\SFOne} &
\resultrow{\textbf{3.07} & 9.32}{\NA}{\NA}{\NA}{\NA}{\NS}{\NA}{66700}{\NA}
& & \textbf{\SFOnePlus} &
\resultrow{\textbf{3.16} & 13.7}{\NA}{\NA}{\NA}{\NA}{\NS}{\NA}{6930}{\NA} \cline{1-3}

\multirow{2}{*}{\textbf{\Adult}} & \multirow{2}{*}{$\mathbf{75 \times 16 \times 5 \times 2 \times 20}$} & \textbf{\AllMarginals} &
\resultrow{\textbf{1.38} & 11.2}{\NA}{\NA}{\NA}{\NA}{\NS}{4.57}{20.5}{\NA}
& & \textbf{\TwoWayMarginals} &
\resultrow{5.30 & 2.11}{\NA}{\NA}{\NA}{\NA}{\NS}{\textbf{2.01}}{155}{\NA} \cline{1-3}

\multirow{2}{*}{\textbf{\CPS}} & \multirow{2}{*}{$\mathbf{100 \times 50 \times 7 \times 4 \times 2}$} & \textbf{\AllRangeMarginals} &
\resultrow{\textbf{1.49} & 421000}{\NA}{\NA}{\NA}{\NA}{\NS}{\NA}{4.74}{\NA}
& & \textbf{\TwoWayRangeMarginals}  &
\resultrow{\textbf{5.79} & 53200}{\NA}{\NA}{\NA}{\NA}{\NS}{\NA}{24.8}{\NA} \hline

\end{tabular}}
\end{table*}


\subsection{Experimental setup} \label{sec:sub:exp_setup}

\vspace{1ex}
\noindent {\bf \em Implementation Details} \quad 
Our Python implementation uses the L-BFGS-B algorithm \cite{byrd1995limited}, as implemented in {\small\texttt{scipy.}} {\small\texttt{optimize}}, as the solver for all optimization routines. Scalability experiments were done on a 4-core Intel i7 3.6GHz processor with 16GB of RAM. The parameter $p$, controlling the size of $p$-Identity strategies for $\optgp$, is set as described in \cref{sec:sub:prog_space}. In experiments not shown, we varied $p$ and found that any choice between $\frac{n}{32}$ and $\frac{n}{8}$ results in nearly the same accuracy.  In experiments, the number of restarts ($S$ in Algorithm 2) is set to 25.  We observed that the distribution of local minima across different random restarts was fairly concentrated, suggesting that far fewer than 25 restarts may be sufficient in practice.

\vspace{1ex}
\noindent{\bf \em Competing techniques} \quad 
We compare \sys against a variety of techniques from the literature.  Some algorithms are specialized to particular settings and we indicate that below.

We consider two baseline algorithms: the Laplace Mechanism (\LMW) and \Identity.  \LMW adds noise directly to each workload query (scaled to the sensitivity)~\cite{inan2016graph,johnson2017elastic}.  \Identity adds noise to the entries of the data vector, then uses the noisy data vector to answer the workload queries. \Identity and \LMW work in both low and high dimensions, on any input workload.

We also consider the two select-measure-reconstruct methods with general search spaces: The Matrix Mechanism (\matrixMech)~\cite{li2015matrix} and the Low Rank Mechanism (\LRM)~\cite{yuan2012low}.  Both of these attempt to find a strategy that minimizes total squared error on the workload queries.

 %

In addition to these general-purpose algorithms, we consider a number of other algorithms designed for low-dim\-en\-sions and specialized for specific workload classes. These include \Privelet~\cite{xiao2011differential}, \HB~\cite{qardaji2013understanding}, \Quadtree~\cite{cormode2012differentially}, and \GreedyH \cite{li2014data}.  These algorithms are all designed to accurately answer range queries:
\Privelet uses a Haar wavelet as the strategy, \HB uses a hierarchical strategy with branching factor that adapts to the domain size, \GreedyH uses a weighted hierarchical strategy, and \Quadtree uses a generalization of the hierarchical strategy to two dimensions.  Of the above methods, \Privelet, \HB, and \Quadtree tend to be quite scalable but have limited search spaces.  \GreedyH solves a non-trivial optimization problem making it less scalable.
For workloads consisting solely of marginals, we also compare against \DataCube~\cite{ding2011differentially}.  \DataCube accepts as input a workload of marginals, and returns a strategy consisting of a different set of marginals that adapts to the input through a greedy heuristic.  

All of the algorithms described so far, with the exception of \LMW, are members of the select-measure-reconstruct paradigm.  We also consider two state-of-the-art algorithms outside of this class: \DAWA~\cite{li2014data} and \PrivBayes~\cite{Zhang2014}.  In 1- or 2-dimensions, the \DAWA algorithm uses some of its privacy budget to detect approximately uniform regions, compresses the domain, reformulates the workload, and then uses the remainder of its privacy budget to run the \GreedyH algorithm described above. \PrivBayes is suitable for multi-dimensional datasets. \PrivBayes first privately fits a Bayes\-ian network on the data and then generates a synthetic dataset by drawing samples from the Bayes\-ian network.  The synthetic data can then be used to answer the workload.  Note that both \DAWA and \PrivBayes have error rates that depend on the input data.

\vspace{1ex}
\noindent {\bf \em Datasets} \quad 
We consider five  datasets, covering low and high dimensional cases.  Most of the algorithms we consider have error rates that only depend on the schema and not the dataset instance. The different schemas we consider have a large impact on the workloads that can be defined over the data and runtime complexity of algorithms.

For 1D and 2D cases, we use representative datasets from the DPBench study~\cite{hay2016principled}, \Patent and BeijingTaxiE (which we call \Taxi).
For higher dimensional cases, we use three datasets, each derived from different Census products. \Census (short for Census of Population and Housing) is the dataset used as a running example throughout the paper and is described in~\cref{sec:usecase}. \Adult is a dataset from the UCI machine learning dataset repository~\cite{adult} with five discrete attributes for age, education, race, sex, and hours-per-week.  \CPS is a dataset released in the March 2000 Population Survey conducted by the Census~\cite{cps}; it has five discrete attributes for income, age, marital status, race, and sex.

For scalability experiments we use synthetic datasets, allowing us to systematically vary the dimensionality and attribute domain sizes.  The runtime of \sys, and the other algorithms we compare against, only depends on the domain size and dimensionality of the data, and not the contents of the data vector, so we use an all-zero data vector.

\vspace{1ex}
\noindent{\bf \em Workloads} \quad 
For the \Census dataset, we use the \SFOne and \SFOnePlus workloads that were introduced in~\cref{sec:usecase} and used as a motivating use case throughout the paper.  For the other datasets, we selected workloads that we believe are representative of typical data analyst interactions.  We also synthesized a few workloads that help illustrate differences in algorithm behavior.


For 1-dimensional datasets, we use three workloads based on range queries: \PrefixOneD, \FixedWidthRange, and \PermutedRange.
\PrefixOneD is $P$, as described in \cref{sec:sub:logical_workloads}, and serves as a compact proxy for all range queries.
The \FixedWidthRange workload contains all range queries that sum $32$ contiguous elements of the domain (i.e., it omits small ranges).
\PermutedRange is a workload consisting of all range queries right-multiplied by a random permutation matrix to randomly shuffle the elements of the domain.  This synthesized workload serves to evaluate whether algorithms can ``recover'' the obscured range query workload.

For 2-dimensional datasets, we use workloads \PrefixTwoD and \PrefixIdentity.  The \PrefixTwoD workload is just the product workload $P \times P$.
The \PrefixIdentity workload is a union of two products: $P \times I$ and $I \times P$.

For higher dimensional datasets, we use a variety of workloads.
We consider multiple workloads based on marginals.  \AllMarginals contains queries for the set of $ 2^d$ marginals (for a dataset of dimension $d$);  \TwoWayMarginals contains queries for the $ \binom{d}{2}$ 2-way marginals; and \ThreeWayMarginals contains queries for the $ \binom{d}{3}$ 3-way marginals.  We also consider a variation on marginals in which range queries are included for numerical attributes (like income and age): \AllRangeMarginals is a marginals-like workload, but the Identity subworkloads are replaced by range query workloads on the numerical attributes, and \TwoWayRangeMarginals is a subset of the previous workload that only contains queries over two dimensions at a time.
\PrefixThreeD is the set of prefix queries along three dimensions ($P \times P\times P$); and \ThreeWayRange is the set of all 3-way range queries.

\vspace{1ex}
\noindent{\bf \em Error measures} \quad 
To compare \sys to other algorithms in terms of accuracy, we report \emph{error ratios}.  Recall from \cref{sec:background} that $Err(W, \algG)$ is the expected total squared error of algorithm $\algG$ for workload $W$, and that for algorithms within the select-measure-reconstruct paradigm, this quantity can be computed in closed form (\cref{def:error}) and is independent of the input data.
Define the \emph{error ratio} between $\algG_{other}$ and \sys on workload $W$ as $Ratio(W, \algG_{other}) = \sqrt{\frac{Err(W,\algG_{other})}{Err(W, \sys)}}$.  Whenever possible, we report analytically computed error ratios (which hold for all settings of $\epsilon$.)
For data-dependent algorithms (DAWA and PrivBayes), the expected error depends on the input data and $\epsilon$, so these are stated in our comparisons.  There is no closed form expression for expected error of a data-dependent algorithm, so we estimate it using average error across 25 random trials.


\subsection{Accuracy comparison} \label{sec:sub:accuracy}

To assess the accuracy of \sys, we considered 11 competing algorithms and a total of 11 workloads, defined over two low-dimensional datasets and three high-dimensional datasets.  Our goal was to empirically compare error of all algorithms in all settings. However, some algorithms are not defined for some of the datasets and workloads; these cases are labeled with $\NA$ in \cref{tab:main}.  For example, a number of algorithms were designed for low-dimensions, while others were designed for high dimensions.  In addition, there are algorithms that are defined for a given dataset/workload, but were infeasible to run; these cases are labeled with $\NS$ in \cref{tab:main}. For example, in theory \matrixMech is applicable to any workload, but it is infeasible to run for the domain sizes we considered. \LRM is also applicable to any workload but is only feasible on medium-sized or smaller domains where the workload and strategy can be represented as a dense matrix. Overall, \sys and the simple baseline methods (\LMW and \Identity) are the only algorithms general and scalable enough to run in all target settings.

\vspace{1ex}
\noindent {\bf \em Findings} \quad
\cref{tab:main} summarizes the results, reported as error ratios to \sys (so that \sys is always 1.0).  It shows that \sys is {\em never outperformed by any competing method} and offers significant improvements, often at least a factor of two and sometimes an order of magnitude.

Even when the workload is low-dimensional range queries, and we compare against a collection of algorithms designed specifically for this workload (e.g. \HB, \Privelet, \GreedyH), \sys is $1.34$ times better than the best algorithm, \HB.
On the \PermutedRange workload, only \sys offers acceptable utility,
since the other algorithms are specifically designed for (unpermuted) range queries, while \sys ad\-apts to a broader class of workloads.

Overall, we see that some algorithms approach the error offered by \sys (ratios close to 1), but, importantly, only for some experimental configurations.  To highlight this, we use bold in the table to indicate the {\em second best} error rate, after \sys.  We find that, depending on the workload and domain size, the second best error rate is achieved by a broad set of algorithms: \Identity, \HB, \Quadtree, \GreedyH are all second-best performers for some workload. This shows that some competing algorithms have specialized capabilities that allow them to perform well in some settings. In contrast, \sys outperforms across all settings, improving error and simplifying the number of algorithms that must be implemented for deployment.

In high dimensions, there are fewer competitors, and \Identity is generally the best alternative to \sys, but the magnitude of improvement by \sys can be as large as 5.79.
Included in \cref{tab:main} are two data-dependent algorithms. \DAWA is defined only for 1D and 2D, but in fact timed out for some low-dimensional workloads. \PrivBayes is designed for high-dimensional data, but does not offer competitive accuracy for these datasets.  Note that the error rates for these methods may differ on different datasets (and for different $\epsilon$ values).

\eat{ 
\paragraph*{Discussion} \mh{improve this discussion; point to specific results} \sys is always the most accurate algorithm, even when the workload is low dimensional range queries, and we compare against a collection of algorithms designed specifically for this task (e.g., \sys is $1.34$ times better than \HB on the \PrefixOneD workload).  The error ratio is even better on the \PrefixTwoD workload, which suggest the magnitude of improvement increasing with dimensionality, which follows theoretically from \cref{thm:err-decomp}.

}

%

\subsection{Scalability comparison} \label{sec:sub:scalability}

\begin{figure*}
\centering
\subcaptionbox{\label{fig:scale1d} N/A: \DataCube}{\includegraphics[width=0.2455 \textwidth]{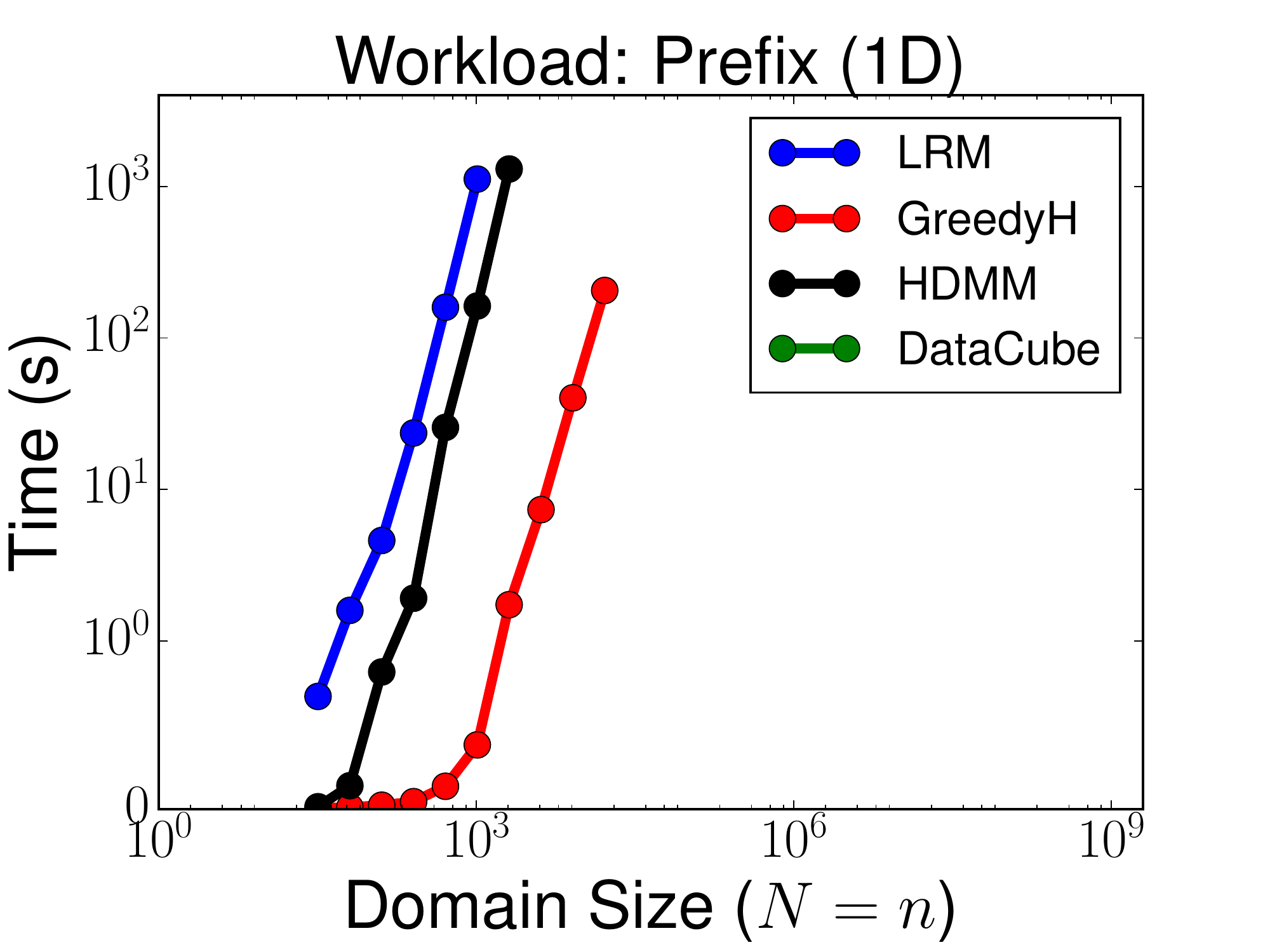}}
\subcaptionbox{\label{fig:scale3d} N/A: \GreedyH, \DataCube}{\includegraphics[width=0.2455 \textwidth]{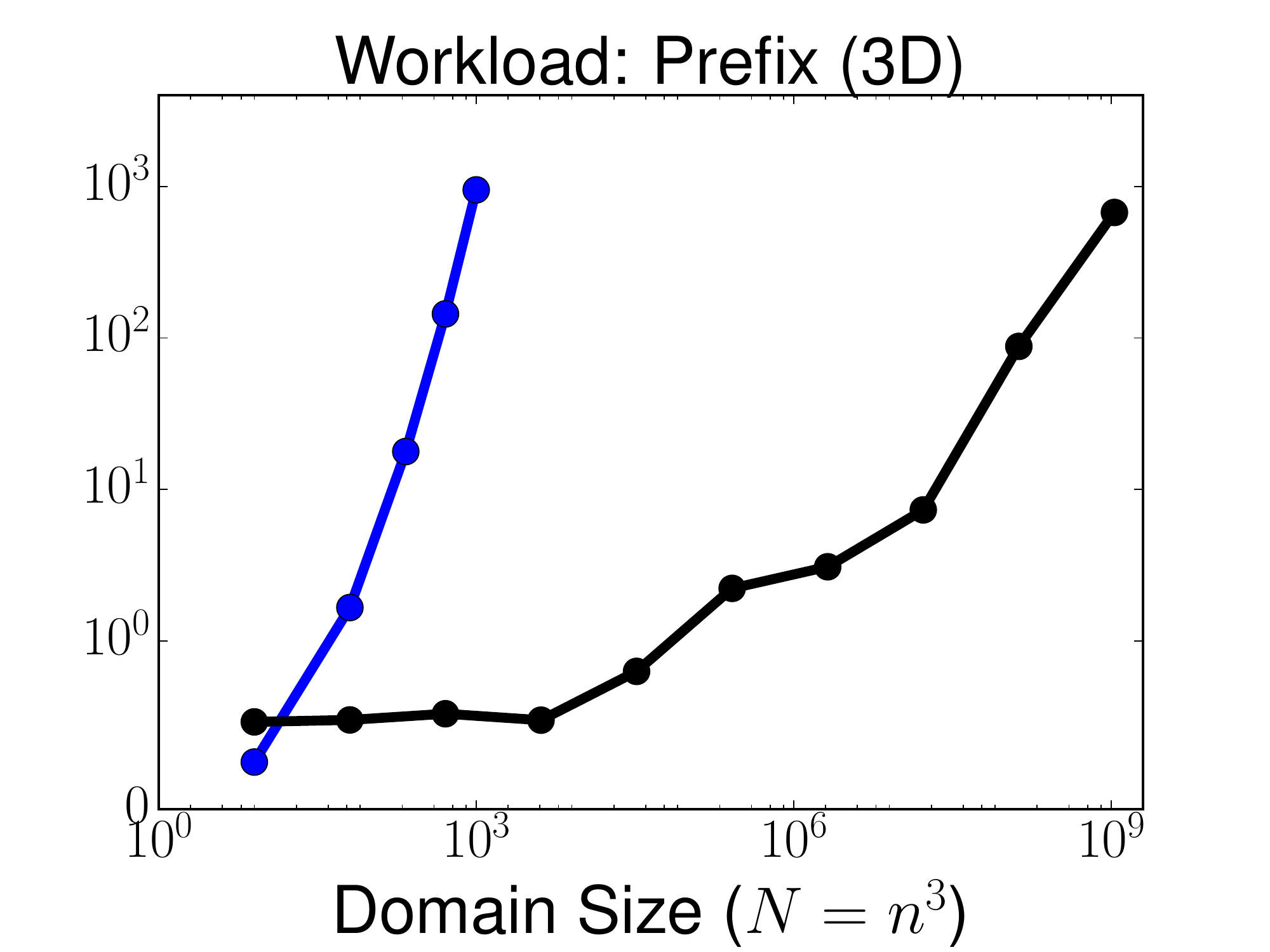}}
\subcaptionbox{\label{fig:scale8d} N/A: \GreedyH}{\includegraphics[width=0.2455 \textwidth]{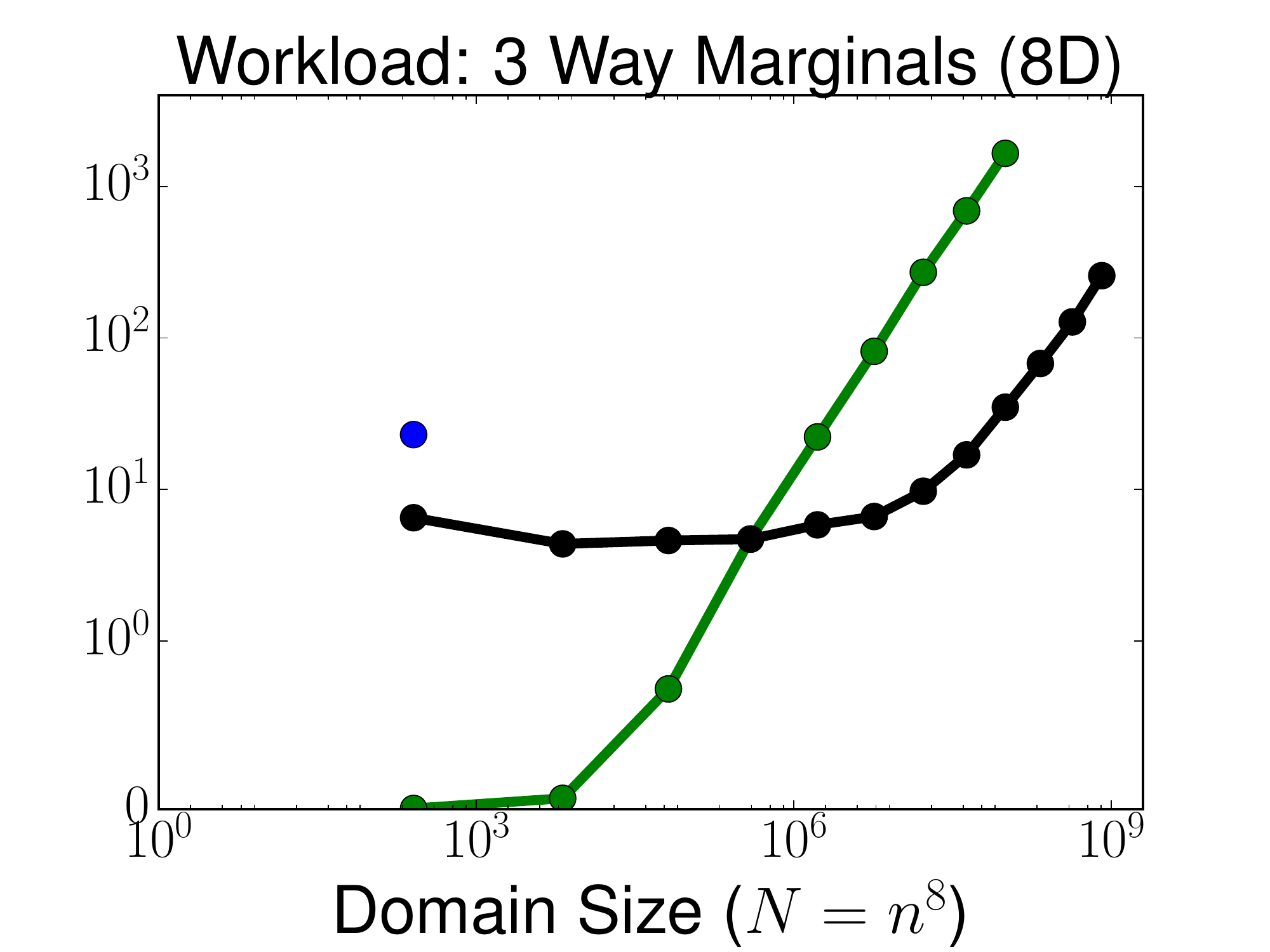}}
\subcaptionbox{\label{fig:scalability_run} Measure+reconstruct}{\includegraphics[width=0.2455 \textwidth]{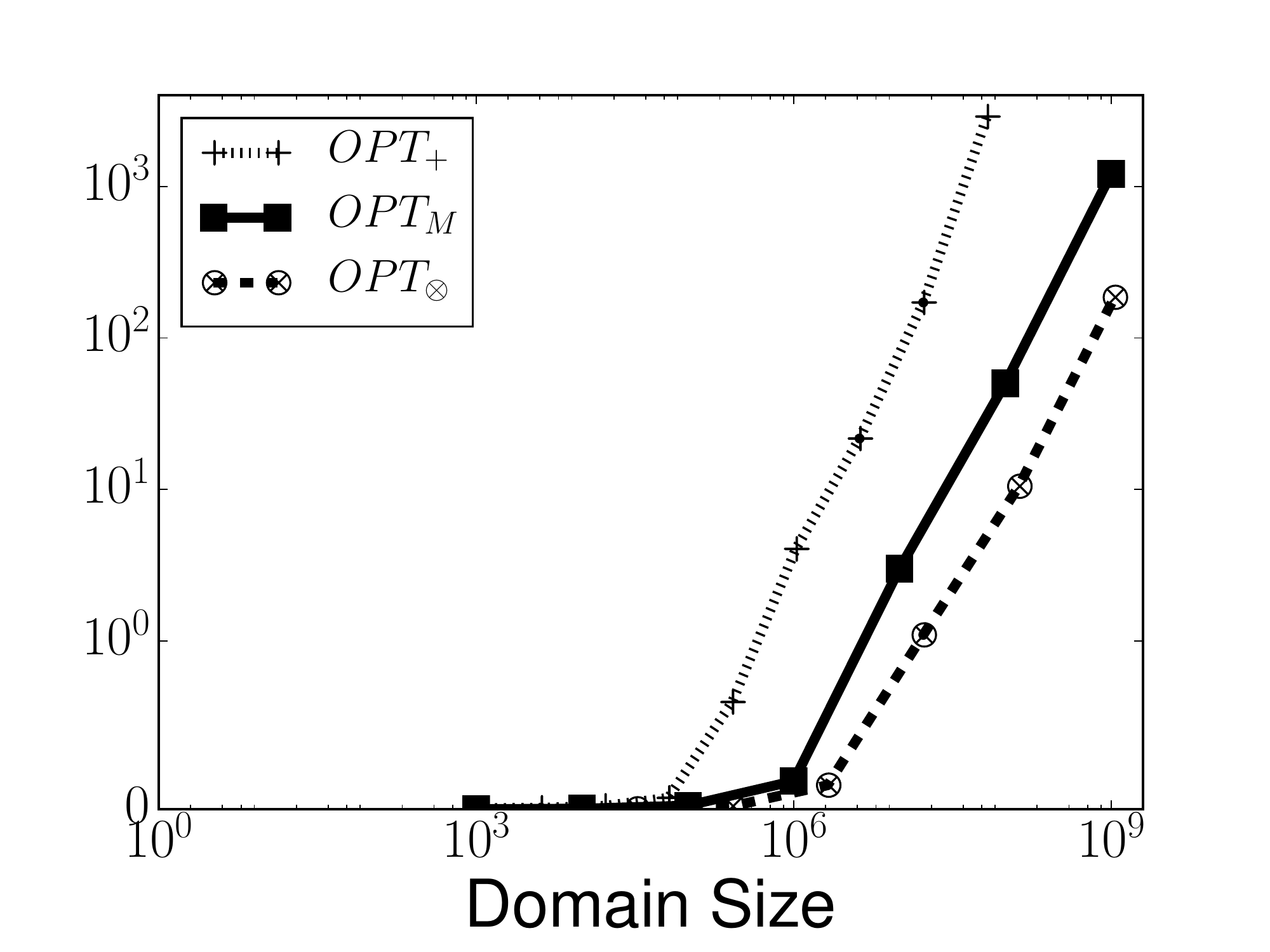}}
\caption{
\label{fig:scalability_main}
(\subref{fig:scale1d})-(\subref{fig:scale8d}) \revision{Runtime comparison on synthetic datasets of increasing domain size (workload and dimensionality varies by plot); (\subref{fig:scalability_run}) Runtime of \sys's measure+reconstruct phase on strategies selected by its $\optk$, $\optkk$ and $\optm$ subroutines.}}
\end{figure*}

Next we evaluate the scalability of \sys by systematically varying the domain size and dimensionality.  We assume throughout that each dimension has a fixed size, $n$, so that $d$-dimensional data has a total domain size of $N=n^d$. We ran each algorithm on increasingly larger datasets until it exhausted memory or reached  a 30-minute timeout.

We compare against the other general-purpose algorithms: \LRM, \GreedyH, and \DataCube.  We omit \matrixMech because it cannot run at these domain sizes and \Identity, \Privelet, \HB, and \Quadtree because (as shown above) they offer sub-optimal error and specialize to narrow workload classes.

\cref{fig:scale1d} shows the scalability of \LRM, \GreedyH, and \sys on the \PrefixOneD workload.  (\DataCube is listed as not applicable (N/A) because it is not defined on non-marginal workloads.)
Since all three of these algorithms require as input an explicitly represented workload in dense matrix form, they are unable to scale beyond $N \approx 10^4$.  On this 1D workload, \sys runs a single instance of $\optgp$, which can be expensive as the domain size grows. \sys is more scalable than \LRM, but less scalable then \GreedyH.  (Recall \sys is run with 25 random restarts; in results not shown, we lower the number of restarts to 1.  At this setting, \sys still achieves higher accuracy than other methods yet achieves lower runtime than \GreedyH.)

\cref{fig:scale3d} shows the scalability of \LRM and \sys on the \PrefixThreeD workload. (\GreedyH is only applicable in 1D; \DataCube remains inapplicable.)  The scalability of \LRM is roughly the same in 3D as it is in 1D because it is determined primarily by the total domain size.  \sys is far more scalable in 3D however, because it solves three smaller optimization problems (for $\optk$) rather than one large problem. The main bottleneck for \sys is measure and reconstruct, not strategy optimization.

\cref{fig:scale8d} shows the scalability of \DataCube and \sys on the \ThreeWayMarginals workload (again, \GreedyH does not apply to high dimensions).  Both \DataCube and \sys ıscale well, with \DataCube scaling to domains as large as $N \approx 10^8$ and \sys scaling to $N \approx 10^9$.  On small domains, \DataCube is faster than \sys, due to \sys's higher up front optimization cost (25 random restarts and multiple optimization programs).  As the domain size grows, measure and reconstruct becomes the bottleneck for both algorithms.  A single data point is shown for \LRM because it did not finish on larger domain sizes ($N \geq 3^8$).

In summary, \sys is most scalable in the multi-dimen\-sional setting, where our implicit representations speed up strategy selection.  The main bottleneck is the measure and reconstruct steps, which we now examine more closely.

Unlike strategy selection, the cost of measure+reconstruct primarily depends on the total domain size (since the data vector must be explicitly represented).  \cref{fig:scalability_run} shows the runtime of measure+reconstruct for synthetic datasets of varying domain sizes.  Since we designed specialized algorithms for each strategy type produced by \sys (as described in \cref{sec:running}), there is one line for each strategy selection subroutine. On strategies produced by $ \optk$ and $\optm$, measure+reconstruct scales to domains as large as $N \approx 10^9$---at which point, the data vector is 4 GB in size (assuming 4 byte floats).  $\optkk$ does not scale as well ($N \approx 10^8$).  This is because computing the pseudo-inverse for $ \optkk $ requires iterative methods, whereas $\optk $ and $\optm$ have closed-form expressions that we exploit.


}

\section{Discussion and conclusions}  \label{sec:conclusion}

\sys is a general and scalable method for privately answering collections of counting queries over high-dimensional data. Because \sys provides state-of-the-art error rates in both low- and high-dimensions, and fully automated strategy selection, we believe it will be broadly useful to algorithm designers. \eatrev{We plan to release source code to enable incorporation of \sys into new applications.}

\revision{\sys is capable of running on multi-dimensional data\-sets with very large domains.  This is primarily enabled by our implicit workload representation in terms of Kronecker products, and our optimization routines for strategy selection that exploit this implicit representation.  We also exploit the structure of the strategies produced by the optimization to efficiently solve the least squares problem.}

\revision{\sys is limited to cases for which it is possible to materialize and manipulate the data vector.  Since we have only investigated a centralized, single-node implementation, it is possible \sys could be scaled to larger data vectors, especially since we have shown that strategy selection is not the bottleneck.  Recent work has shown that standard operations on large matrices can be parallelized \cite{Xiang14Scalable}, however the decomposed structure of our strategies should lead to even faster specialized parallel solutions.
Ultimately, for very large domains, factoring the data (as PrivBayes does) may be unavoidable. \sys still has a role to play, however, since it can be used to optimize queries over the factored subsets of the data.}


\revision{As noted previously, HDMM optimizes for absolute error and is not applicable to optimizing relative error, which is data dependent.  Nevertheless, by weighting the workload queries (e.g. inversely with their $L_1$-norm) we can approximately optimize relative error, at least for datasets whose data vectors are close to uniform.  This approach could be extended to reflect a user's assumptions or guesses about the input data.  Future work could also integrate HDMM measurement with techniques like iReduct \cite{Xiao11iReduct:} which perform adaptive measurement to target relative error.}

While \sys produces the best known strategies for a variety of workloads, we do not know how close to optimal its solutions are.  There are asymptotic lower bounds on error in the literature \cite{Gupta12Iterative,nikolov2013geometry,hardt2010geometry}, but it is not clear how to use them on a concrete case given hidden constant factors.  Li et al \cite{Li13Optimal} provided a precise lower bound in terms of the spectral properties of $\W$, but it is not clear how to compute it on our large workload matrices and it is often a very loose lower bound under $\epsilon$-differential privacy.  
\eatrev{Under $(\epsilon, \delta)$-differential privacy the bound is tighter, and in addition, the strategy optimization problem is more tractable \cite{li2010optimizing,li2012adaptive,li2015matrix}. Yuan et al. \cite{yuan2016convex} recently found a convex optimization solution that could serve the role of $\optgp$. In conjunction with $\optk$ and $\optkk$, it could be the basis of a version of \sys adapted to $(\epsilon, \delta)$-differential privacy.}

\eatrev{Recall that our techniques support {\em arbitrary} predicates on individual attributes that are combined using conjunctions.  We showed that more complex conditions on multiple attributes can be handled by merging attributes (see \cref{ex:imp_vec}). The limitation is that if a merged attribute gets too big, it becomes a challenge to optimize with $\optgp$.


	A promising alternative relies on {\em workload approximation}.  Given an input workload $\Wlog$, it is possible to perform strategy selection with a surrogate workload, $\Wlog'$, as input, still using the output strategy to answer $\Wlog$.  This changes the objective function, but if $\Wlog'$ approximates $\Wlog$ well, it typically has little effect on solution quality.  
	If $\Wlog$ includes disjunctions, we could construct an approximate surrogate workload $\Wlog'$ by replacing disjunctions by one or more conjunctive queries. Another application of workload approximation improves efficiency: we may find an approximate workload consisting of fewer products (i.e. smaller $k$).  Initial results suggest that our problems are highly tolerant of optimization on surrogate workloads.}  

\vspace{1ex}
{\footnotesize
\noindent\textbf{Acknowledgements:}  This work was supported by the National Science Foundation under grants 1253327, 1408982, 1409125, 1443014, 1421325, and 1409143; and by DARPA and SPAWAR under contract N66001-15-C-4067. The U.S. Government is authorized to reproduce and distribute reprints for Governmental purposes not withstanding any copyright notation thereon. The views, opinions, and/or findings expressed are those of the author(s) and should not be interpreted as representing the official views or policies of the Department of Defense or the U.S. Government. \par
}

\clearpage
\interlinepenalty=10000
\bibliographystyle{abbrv}
\bibliography{bib/refs}

\begin{thebibliography}{10}

\bibitem{cps}
Current population survey.
\newblock Available at:
  {https://www2.census.gov/programs-surveys/cps/techdocs/cpsmar00.pdf}, March
  2000.

\bibitem{Acs2012compression}
G.~{\'A}cs, C.~Castelluccia, and R.~Chen.
\newblock Differentially private histogram publishing through lossy
  compression.
\newblock In {\em ICDM}, pages 1--10, 2012.

\bibitem{barak2007privacy}
B.~Barak, K.~Chaudhuri, C.~Dwork, S.~Kale, F.~McSherry, and K.~Talwar.
\newblock Privacy, accuracy, and consistency too: a holistic solution to
  contingency table release.
\newblock In {\em Proceedings of the twenty-sixth ACM SIGMOD-SIGACT-SIGART
  symposium on Principles of database systems}, pages 273--282. ACM, 2007.

\bibitem{bhaskara2012unconditional}
A.~Bhaskara, D.~Dadush, R.~Krishnaswamy, and K.~Talwar.
\newblock Unconditional differentially private mechanisms for linear queries.
\newblock In {\em Proceedings of the Forty-fourth Annual ACM Symposium on
  Theory of Computing}, STOC '12, pages 1269--1284, New York, NY, USA, 2012.
  ACM.

\bibitem{byrd1995limited}
R.~H. Byrd, P.~Lu, J.~Nocedal, and C.~Zhu.
\newblock A limited memory algorithm for bound constrained optimization.
\newblock {\em SIAM Journal on Scientific Computing}, 16(5):1190--1208, 1995.

\bibitem{census-sf1}
2010 {C}ensus {S}ummary {F}ile 1, {C}ensus of {P}opulation and {H}ousing.
\newblock Available at {\tt https://www.census.gov/prod/cen2010/doc/sf1.pdf},
  2012.

\bibitem{census-url}
Census scientific advisory committee meeting.
\newblock www.census.gov/about/cac/sac/meetings/2017-09-meeting.html, September
  2017.

\bibitem{cormode2012differentially}
G.~Cormode, C.~Procopiuc, D.~Srivastava, E.~Shen, and T.~Yu.
\newblock Differentially private spatial decompositions.
\newblock In {\em Data engineering (ICDE), 2012 IEEE 28th international
  conference on}, pages 20--31. IEEE, 2012.

\bibitem{adult}
D.~Dheeru and E.~Karra~Taniskidou.
\newblock {UCI} machine learning repository, 2017.

\bibitem{ding2011differentially}
B.~Ding, M.~Winslett, J.~Han, and Z.~Li.
\newblock Differentially private data cubes: optimizing noise sources and
  consistency.
\newblock In {\em Proceedings of the 2011 ACM SIGMOD International Conference
  on Management of data}, pages 217--228. ACM, 2011.

\bibitem{dwork2006calibrating}
C.~Dwork, F.~M.~K. Nissim, and A.~Smith.
\newblock Calibrating noise to sensitivity in private data analysis.
\newblock In {\em TCC}, pages 265--284, 2006.

\bibitem{Dwork14Algorithmic}
C.~Dwork and A.~Roth.
\newblock {\em The Algorithmic Foundations of Differential Privacy}.
\newblock Found. and Trends in Theoretical Computer Science, 2014.

\bibitem{ebadi2015personal}
H.~Ebadi, D.~Sands, and G.~Schneider.
\newblock Differential privacy: Now it's getting personal.
\newblock In {\em Proceedings of the 42Nd Annual ACM SIGPLAN-SIGACT Symposium
  on Principles of Programming Languages}, POPL '15, pages 69--81, New York,
  NY, USA, 2015. ACM.

\bibitem{fong2011lsmr}
D.~C.-L. Fong and M.~Saunders.
\newblock Lsmr: An iterative algorithm for sparse least-squares problems.
\newblock {\em SIAM Journal on Scientific Computing}, 33(5):2950--2971, 2011.

\bibitem{Gupta12Iterative}
A.~Gupta, A.~Roth, and J.~Ullman.
\newblock Iterative constructions and private data release.
\newblock In {\em Proceedings of the 9th International Conference on Theory of
  Cryptography}, TCC'12, pages 339--356, Berlin, Heidelberg, 2012.
  Springer-Verlag.

\bibitem{hager1989updating}
W.~W. Hager.
\newblock Updating the inverse of a matrix.
\newblock {\em SIAM review}, 31(2):221--239, 1989.

\bibitem{sigmod:haney17}
S.~Haney, A.~Machanavajjhala, M.~Kutzbach, M.~Graham, J.~Abowd, and
  L.~Vilhuber.
\newblock Utility cost of formal privacy for releasing national
  employer-employee statistics.
\newblock In {\em ACM SIGMOD}, 2017.

\bibitem{hardt2010geometry}
M.~Hardt and K.~Talwar.
\newblock On the geometry of differential privacy.
\newblock In {\em Proceedings of the Forty-second ACM Symposium on Theory of
  Computing}, STOC '10, pages 705--714, New York, NY, USA, 2010. ACM.

\bibitem{hay2016principled}
M.~Hay, A.~Machanavajjhala, G.~Miklau, Y.~Chen, and D.~Zhang.
\newblock Principled evaluation of differentially private algorithms using
  dpbench.
\newblock In {\em Proceedings of the 2016 International Conference on
  Management of Data}, pages 139--154. ACM, 2016.

\bibitem{hay2010boosting}
M.~Hay, V.~Rastogi, G.~Miklau, and D.~Suciu.
\newblock Boosting the accuracy of differentially private histograms through
  consistency.
\newblock {\em PVLDB}, 3(1-2):1021--1032, 2010.

\bibitem{hcupnet}
{HCUPnet: Healthcare Cost and Utilization Project}.
\newblock Available at {\tt https://hcupnet.ahrq.gov/}.

\bibitem{inan2016graph}
A.~Inan, M.~E. Gursoy, E.~Esmerdag, and Y.~Saygin.
\newblock Graph-based modelling of query sets for differential privacy.
\newblock In {\em Proceedings of the 28th International Conference on
  Scientific and Statistical Database Management}, SSDBM '16, pages 3:1--3:10,
  New York, NY, USA, 2016. ACM.

\bibitem{johnson2017elastic}
N.~Johnson, J.~P. Near, and D.~Song.
\newblock Towards practical differential privacy for sql queries.
\newblock {\em PVLDB}, 11(5):526--539, 2018.

\bibitem{tods:Kifer14}
D.~Kifer and A.~Machanavajjhala.
\newblock Pufferfish: A framework for mathematical privacy definitions.
\newblock {\em ACM Transactions on Database Systems (TODS)}, 39(1):1--36, 2014.

\bibitem{li2014data}
C.~Li, M.~Hay, G.~Miklau, and Y.~Wang.
\newblock A data-and workload-aware algorithm for range queries under
  differential privacy.
\newblock {\em PVLDB}, 7(5):341--352, 2014.

\bibitem{li2010optimizing}
C.~Li, M.~Hay, V.~Rastogi, G.~Miklau, and A.~McGregor.
\newblock Optimizing linear counting queries under differential privacy.
\newblock In {\em Proceedings of the twenty-ninth ACM SIGMOD-SIGACT-SIGART
  symposium on Principles of database systems}, pages 123--134. ACM, 2010.

\bibitem{li2012adaptive}
C.~Li and G.~Miklau.
\newblock An adaptive mechanism for accurate query answering under differential
  privacy.
\newblock {\em PVLDB}, 5(6):514--525, 2012.

\bibitem{Li13Optimal}
C.~Li and G.~Miklau.
\newblock Optimal error of query sets under the differentially-private matrix
  mechanism.
\newblock In {\em ICDT}, 2013.

\bibitem{li2015matrix}
C.~Li, G.~Miklau, M.~Hay, A.~McGregor, and V.~Rastogi.
\newblock The matrix mechanism: optimizing linear counting queries under
  differential privacy.
\newblock {\em The VLDB Journal}, 24(6):757--781, 2015.

\bibitem{onthemap:icde08}
A.~Machanavajjhala, D.~Kifer, J.~M. Abowd, J.~Gehrke, and L.~Vilhuber.
\newblock Privacy: Theory meets practice on the map.
\newblock In {\em ICDE}, pages 277--286, 2008.

\bibitem{mcsherry2009pinq}
F.~D. McSherry.
\newblock Privacy integrated queries: An extensible platform for
  privacy-preserving data analysis.
\newblock In {\em Proceedings of the 2009 ACM SIGMOD International Conference
  on Management of Data}, SIGMOD '09, pages 19--30, New York, NY, USA, 2009.
  ACM.

\bibitem{nikolov2013geometry}
A.~Nikolov, K.~Talwar, and L.~Zhang.
\newblock The geometry of differential privacy: The sparse and approximate
  cases.
\newblock In {\em Proceedings of the Forty-fifth Annual ACM Symposium on Theory
  of Computing}, STOC '13, pages 351--360, New York, NY, USA, 2013. ACM.

\bibitem{onthemap}
{OnTheMap Web Tool}.
\newblock Available at {\tt http://onthemap.ces.census.gov/}.

\bibitem{petersen2008matrix}
K.~B. Petersen, M.~S. Pedersen, et~al.
\newblock The matrix cookbook.
\newblock {\em Technical University of Denmark}, 7:15, 2008.

\bibitem{qardaji2013differentially}
W.~Qardaji, W.~Yang, and N.~Li.
\newblock Differentially private grids for geospatial data.
\newblock In {\em Intl. Conference on Data Engineering (ICDE)}, pages 757--768.
  IEEE, 2013.

\bibitem{qardaji2013understanding}
W.~Qardaji, W.~Yang, and N.~Li.
\newblock Understanding hierarchical methods for differentially private
  histograms.
\newblock {\em PVLDB}, 6(14):1954--1965, 2013.

\bibitem{qardaji2014priview}
W.~Qardaji, W.~Yang, and N.~Li.
\newblock Priview: practical differentially private release of marginal
  contingency tables.
\newblock In {\em Proceedings of the 2014 ACM SIGMOD international conference
  on Management of data}, pages 1435--1446. ACM, 2014.

\bibitem{vaidya2013hcupnet}
J.~Vaidya, B.~Shafiq, X.~Jiang, and L.~Ohno-Machado.
\newblock Identifying inference attacks against healthcare data repositories.
\newblock {\em AMIA Jt Summits Transl Sci Proc}, 2013, 2013.

\bibitem{van2000ubiquitous}
C.~F. Van~Loan.
\newblock The ubiquitous kronecker product.
\newblock {\em Journal of computational and applied mathematics},
  123(1):85--100, 2000.

\bibitem{Xiang14Scalable}
J.~Xiang, H.~Meng, and A.~Aboulnaga.
\newblock Scalable matrix inversion using mapreduce.
\newblock In {\em High-performance Parallel and Distributed Computing}, HPDC
  '14, 2014.

\bibitem{Xiao11iReduct:}
X.~Xiao, G.~Bender, M.~Hay, and J.~Gehrke.
\newblock ireduct: Differential privacy with reduced relative errors.
\newblock In {\em SIGMOD}, 2011.

\bibitem{xiao2008output}
X.~Xiao and Y.~Tao.
\newblock Output perturbation with query relaxation.
\newblock {\em PVLDB}, 1(1):857--869, Aug. 2008.

\bibitem{xiao2011differential}
X.~Xiao, G.~Wang, and J.~Gehrke.
\newblock Differential privacy via wavelet transforms.
\newblock {\em IEEE Transactions on Knowledge and Data Engineering},
  23(8):1200--1214, 2011.

\bibitem{xiao2014dpcube}
Y.~Xiao, L.~Xiong, L.~Fan, S.~Goryczka, and H.~Li.
\newblock {DPCube}: Differentially private histogram release through
  multidimensional partitioning.
\newblock {\em Transactions of Data Privacy}, 7(3), 2014.

\bibitem{xu12histogram}
J.~Xu, Z.~Zhang, X.~Xiao, Y.~Yang, and G.~Yu.
\newblock Differentially private histogram publication.
\newblock In {\em Data Engineering (ICDE), 2012 IEEE 28th International
  Conference on}, pages 32--43, 2012.

\bibitem{xu2013differential}
J.~Xu, Z.~Zhang, X.~Xiao, Y.~Yang, G.~Yu, and M.~Winslett.
\newblock Differentially private histogram publication.
\newblock {\em The VLDB Journal}, pages 1--26, 2013.

\bibitem{Yaroslavtsev13Accurate}
G.~Yaroslavtsev, G.~Cormode, C.~M. Procopiuc, and D.~Srivastava.
\newblock Accurate and efficient private release of datacubes and contingency
  tables.
\newblock In {\em ICDE}, 2013.

\bibitem{yuan2016convex}
G.~Yuan, Y.~Yang, Z.~Zhang, and Z.~Hao.
\newblock Convex optimization for linear query processing under approximate
  differential privacy.
\newblock In {\em Proceedings of the 22nd ACM SIGKDD International Conference
  on Knowledge Discovery and Data Mining}, pages 2005--2014. ACM, 2016.

\bibitem{yuan2012low}
G.~Yuan, Z.~Zhang, M.~Winslett, X.~Xiao, Y.~Yang, and Z.~Hao.
\newblock Low-rank mechanism: optimizing batch queries under differential
  privacy.
\newblock {\em PVLDB}, 5(11):1352--1363, 2012.

\bibitem{Zhang2014}
J.~Zhang, G.~Cormode, C.~M. Procopiuc, D.~Srivastava, and X.~Xiao.
\newblock Privbayes: Private data release via bayesian networks.
\newblock {\em ACM Transactions on Database Systems (TODS)}, 42(4):25, 2017.

\bibitem{zhang16privtree}
J.~Zhang, X.~Xiao, and X.~Xie.
\newblock Privtree: A differentially private algorithm for hierarchical
  decompositions.
\newblock In {\em SIGMOD}, 2016.

\bibitem{zhangtowards}
X.~Zhang, R.~Chen, J.~Xu, X.~Meng, and Y.~Xie.
\newblock Towards accurate histogram publication under differential privacy.
\newblock In {\em SDM}, 2014.

\end{thebibliography}


\clearpage
\appendix
\section{Proofs and Technical Details}

\subsection{Proof of Theorem 2}

In this section, we prove that any logical product workload can be represented in matrix form as a Kronecker product, as long as the data vector is organized in a particular order.

For now assume the domain is two dimensional, with relational schema $ R(A, B) $.  It is straightforward to extend to higher dimensional domains using induction.  Recall from~\cref{sec:background} that the data vector $\x$ is indexed by tuples in the domain $t \in dom(R)$ such that $\x_t$ counts the number of occurrences of $t$ in the data.  The data matrix $\X$, defined below, is an alternative representation of this that exposes the two-dimensional structure of the data. 

\begin{definition}[Data Matrix]
The data matrix $\X$ is indexed by tuples $ (a,b) \in dom(A) \times dom(B) $ such that $ \X_{ab} $ counts the number of occurrences of $ (a,b) $ in the data.  
\end{definition}

It is easy to go back and forth between the two representations of the data through flattening and reshaping.  In fact, the data vector $\x$ corresponding to a data matrix $\X$ is simply $ \x = flat(\X)$ where $flat$ flattens a matrix into a column vector by vertically stacking the transposed rows of the matrix.  

As shown in~\cref{prop:conjunct}, predicate counting queries with conjunctive predicates can naturally be expressed in terms of the data matrix $\X$.  

\begin{proposition} \label{prop:conjunct}
For two predicate counting queries \\ $ \phi : dom(A) \rightarrow \set{0,1} $ and $ \psi : dom(B) \rightarrow \set{0,1} $, 
$$ vec(\phi \wedge \psi) \x = vec(\phi) \X vec(\psi)^T $$
\end{proposition}

\begin{proof}
\begin{align*}
vec(\phi) \X vec(\psi)^T &= \sum_{a \in dom(A)} \sum_{b \in dom(B)} \phi(a) \psi(b) \X_{ab} \\
&= \sum_{a \in dom(A)} \sum_{b \in dom(B)}  (\phi \wedge \psi)(a,b) \X_{ab} \\
&= \sum_{t \in dom(R)} (\phi \wedge \psi)(t) \x_t  \\
&= vec(\phi \wedge \psi) \x
\end{align*}
\end{proof}

Additionally,~\cref{prop:conjunct} is useful to show that product workloads (\cref{def:product-workload}) can be represented in matrix form as a Kronecker product, as in~\cref{prop:kron}.

\begin{proposition} \label{prop:kron}
For two predicate counting query sets $\Phi = [\phi_1 \dots \phi_p]_A$ and $\Psi = [\psi_1 \dots \psi_r]_B$, 
$$vec(\Phi \times \Psi) = vec(\Phi) \otimes vec(\Psi) $$
\end{proposition}

\begin{proof}
Let $\x$ be any data vector and let $\X$ be the corresponding data matrix.  
We will show that $ vec(\Phi \times \Psi) \x = (vec(\Phi) \otimes vec(\Psi)) \x$ for all $\x$.  We know $ (\A \otimes \B) \x = flat( \A \X \B^T ) $ by the matrix equations identity for Kronecker products.  Thus,
$$ (vec(\Phi) \otimes vec(\Psi)) \x = flat( vec(\Phi) \X vec(\Psi)^T ) $$
where 
$$ [ vec(\Phi) \X vec(\Psi)^T ]_{ij} = vec(\phi_i) \X vec(\psi_j)^T $$
By~\cref{prop:conjunct}, this is equvialent to $ vec(\phi_i \wedge \psi_j) \x $.  Since this gives us $ vec(\phi_i \wedge \psi_j) \x $ for each $ \phi_i \in \Phi $ and each $\psi_j \in \Psi$, it computes the answers to all the product queries $\Phi \times \Psi$.
\end{proof}


\subsection{Proof of Formula~\ref{eq:grad}}

In \cref{sec:optimization}, we made the following claim:
\begin{equation}
\begin{aligned}
\frac{\partial C}{\partial \A} &= -2 \A (\A^T \A)^+ (\W^T \W) (\A^T \A)^+
\end{aligned} \tag{\ref{eq:grad}}
\end{equation}
First note that $C(\A)$ can be expressed as $tr[\X^+ \Y]$ where $\X = \A^T \A$ and $\Y = \W^T \W$  From matrix calculus~\cite{petersen2008matrix}, we know that $ \frac{\partial C}{\partial \X} = -\X^{-1} \Y \X^{-1} $ if $ \X $ is invertible, and more generally
\begin{align*}
\frac{\partial C}{\partial \X} = &-\X^+ \Y \X^+ \\&+ (\X^+ \X^{+T}) \Y (\I - \X \X^+) \\&+ (\I - \X^+ \X) \Y (\X^{+T} \X^+)
\end{align*}
which simplifies to $ \frac{\partial C}{\partial \X} = -\X^+ \Y \X^+ $ when $\A$ supports $\W$.  Equivalently, 
$$ \frac{\partial C}{\partial \X} = -(\A^T \A)^+ (\W^T \W) (\A^T \A)^+ $$
%


The gradient with respect to $\A$ can be written in terms of the gradient with respect to $\X$ using the chain rule:
\begin{align*}
\frac{\partial C}{\partial \A} &= 2 \A \frac{\partial C}{\partial \X} \\
&= -2 \A (\A^T \A)^+ (\W^T \W) (\A^T \A)^+
\end{align*}
Thus, the claim holds (as long as $\A$ supports $\W$).  

Note that in order to apply gradient based optimization with respect to a parameterized strategy, such as the p-Identity strategies $\A(\vect{\Theta})$ discussed in~\cref{sec:sub:generalpurpose}, it is necessary to apply the chain rule again:
$$ \frac{\partial C}{\partial \vect{\Theta}_{kl}} = \sum_{i,j} \frac{\partial C}{\partial \A_{ij}} \frac{\partial \A_{ij}}{\partial \vect{\Theta}_{kl}} $$

The computational cost of this step is small compared to the cost of calculating $ \frac{\partial C}{\partial \A} $.

\subsection{Analysis of $\optgp$}

Below is the proof of \cref{thm:optgp} showing the complexity of the objective and its gradient for $\optgp$.

\thmoptgp

\begin{proof}
Assume $\W^T \W$ has been precomputed
and now express $ \A^T \A $ in terms of $\matr{\Theta}$ and $\D$:
$$ \A^T \A = \D^T \D + \D^T \matr{\Theta}^T \matr{\Theta} \D = \D[\I_n + \matr{\Theta}^T \matr{\Theta}] \D $$
Applying the identity $ (\X \matr{Y})^{-1} = \matr{Y}^{-1} \X^{-1} $ together with the Woodbury identity~\cite{hager1989updating} yields an expression for the inverse:
\begin{align*}
(\A^T \A)^{-1} &= \D^{-1} [\I_n + \matr{\Theta}^T \matr{\Theta}]^{-1} \D^{-1} \\
&= \D^{-1} [\I_n - \matr{\Theta}^T (\I_p + \matr{\Theta} \matr{\Theta}^T)^{-1} \matr{\Theta}] \D^{-1}
\end{align*}

We can compute $ (\A^T \A)^{-1} (\W^T \W) $ in $O(n^2 p)$ time by evaluating the following expression from right to left:
\begin{align*}
(\A^T \A)^{-1} (\W^T \W) &= \D^{-2} (\W^T \W) \\ &- \D^{-1} \matr{\Theta}^T (\I_p + \matr{\Theta} \matr{\Theta}^T)^{-1} \matr{\Theta} \D^{-1} (\W^T \W)
\end{align*}

By carefully looking at the dimensionality of the intermediate matrices that arise from carrying out the matrix multiplications from right-to-left, we see that the most expensive operation is the matrix-matrix product between an $ n \times p$ matrix and a $p \times n$ matrix, which takes $O(n^2 p)$ time.  The inverse $ (\I_p + \matr{\Theta} \matr{\Theta}^T)^{-1} $ takes $O(p^3)$ time and the operations involving $\D$ take $O(n^2)$ time since it is a diagonal matrix.

The result still holds even if $\W^T \W$ is replaced with an arbitrary $n \times n$ matrix, so $ \X = (\A^T \A)^{-1} (\W^T \W) (\A^T \A)^{-1} $ can be computed in $O(n^2 p)$ time as well.  The gradient is $ -2 \A \X $ whose components can be calculated separately as $ -2 \D \X $ and $ -2 \matr{\Theta} \X $.  $ -2 \D \X $ takes $O(n^2)$ time and $ \matr{\Theta} \X$  takes $O(n^2 p)$ time, so the overall cost of computing the gradient is $O(n^2 p)$.
\end{proof}

\subsection{Marginals parameterization}

We now discuss in more detail how we solve Problem \ref{prob:marginal} from \cref{sec:sub:marginals}.  In particular, we show how to efficiently find the pseudo inverse of $\MM(\vect{\theta})$, which is required to evaluate the objective function.  

It is useful to formally define a correspondence between the integers $ [2^d] = \set{0, \dots, 2^d-1} $ and the query matrices for each marginal.  We therefore define $ \CC : [2^d] \rightarrow \mathbb{R}^{N \times N} $ as follows:
$$ \CC(a) = \bigotimes_{i=1}^d [\matr{1} (a_i = 0) + \I (a_i = 1)] $$
where $(a_i = 0)$ and $(a_i = 1)$ are indicator functions on the $i^{th}$ bit of the binary representation of $a$ and $\matr{1} = \T^T \T$ is the $n_i \times n_i$ matrix of ones.

We also define $\GG : \mathbb{R}^{2^d} \rightarrow \mathbb{R}^{N \times N}$ as follows:
$$ \GG(v) = \sum_{a \in [2^d]} v_a \CC(a) $$
and note that $ \AA^T \AA = \GG(\vect{\theta}^2)$ for a strategy $\AA = \MM(\vect{\theta})$. 

\begin{example}
For a 3 dimensional domain the query matrix for a 2 way marginal can be expressed as $ \QQ = \I \otimes \T \otimes \I $, and $\QQ^T \QQ = \I \otimes \matr{1} \otimes \I = \CC(101_2) = \CC(5) $.
\end{example}

\cref{prop:marginals} shows that matrices $\CC(a)$ and $\CC(b)$ interact nicely under matrix multiplication.

\begin{proposition} \label{prop:marginals}
For any $a, b \in [2^d] $, $$\CC(a) \CC(b) = C(a | b) \CC(a \& b)$$ where $ a | b $ denotes ``bitwise or'', $ a \& b $ denotes ``bitwise and'', and $C(k) = \prod_{i = 1}^d [n_i (k_i=0) + 1 (k_i=1)] $.
\end{proposition}

\begin{proof}
First observe how the matrices $\I$ and $\matr{1}$ interact under matrix multiplication:
\begin{align*}
\I \I = \I && \I \matr{1} = \matr{1} && \matr{1} \I = \matr{1} && \matr{1} \matr{1} = n_i \matr{1} \\
\end{align*}
Now consider the product $\CC(a) \CC(b)$ which is simplified using Kronecker product identities, logical rules, and bitwise manipulation.
\begin{align*}
= &\bigotimes_{i=1}^d [\matr{1} (a_i = 0) + \I (a_i = 1)] [\matr{1} (b_i = 0) + \I (b_i = 1)]  \\
= &\prod_{i=1}^d [n_i (a_i = 0 \text{ and } b_i = 0) + 1 (a_i = 1 \text{ or } b_i = 1)] \\
  &\bigotimes_{i=1}^d [\matr{1} (a_i = 0 \text{ or } b_i = 0) + \I (a_i = 1 \text{ and } b_i = 1)] \\
= &\prod_{i=1}^d [ n_i ( (a | b)_i = 0) + 1 ( (a | b)_i = 1) ] \\
  &\bigotimes_{i=1}^d [\matr{1} ( (a \& b)_i = 0) + \I ( (a \& b)_i = 1)] \\
= &C(a|b) \CC(a \& b)
\end{align*}
\end{proof}

From \cref{prop:marginals}, it follows that $ \set{c \GG(v) \mid c \in \mathbb{R}, v \in \mathbb{R}^{2^d}}$ is closed under matrix multiplication.  In fact, the relationship is linear: for fixed $u$, the product $ \GG(u) \GG(v) $ is linear in $v$.

\begin{proposition} \label{prop:marginals2}
For any $u, v \in \mathbb{R}^{2^d}$, $$\GG(u) \GG(v) = \GG(\X(u) v)$$ where $\X(u)$ is a triangular matrix that depends on $u$.
\end{proposition}

\begin{proof}
Let $u, v \in \mathbb{R}^{2^d}$ and consider the following product:
\begin{align*}
\GG(u) \GG(v) &= \Big( \sum_a u_a \CC(a) \Big) \Big( \sum_b v_b \CC(b) \Big) \\
&= \sum_{a,b} u_a v_b \CC(a) \CC(b) \\
&= \sum_{a,b} u_a v_b C(a | b) \CC(a \& b) \\
\end{align*}
Observe that $ \GG(u) \GG(v) = \GG(w) $ where
$$ w_k = \sum_{a \& b = k} u_a v_b C(a | b) $$
The relationship between $w$ and $v$ is clearly linear, and by carefully inspecting the expression one can see that $ w = \X(u) v $ where $ \X(u)[k,b] = \sum_{a : a \& b = k} u_a C(a | b) $.  $\X(u)$ is an upper triangular matrix because $k = a\&b$ and $ a \& b \leq b$ for all $a$.
\end{proof}

We can use \cref{prop:marginals2} to efficiently multiply two matrices of this form in the compact representation, and we can also apply it to find the inverse or a generalized inverse of a matrix of this form.  In particular, if $v$ is a solution to the linear system $\X(u) v = z$ where $\GG(z) = \I $ then $ \GG(u)^{-1} = \GG(v) $.  A similar linear system can be solved to find a generalized inverse or the pseudo inverse if $ \GG(u) $ is not invertible.

This gives us all the machinery we need to efficiently compute the pseudo inverse, and consequently the objective function and its gradient.  As a result, we can apply gradient-based techniques to optimize the strategy of marginals.

\begin{table*}
\centering
\subcaptionbox{\label{tbl:marg-1d} 1D workloads}{
\resizebox{.48\textwidth}{!}{
\begin{tabular}{cc|SSSS|S}
\textbf{Workload}                        & \textbf{Domain} & \textbf{Identity} & \textbf{Wavelet} & \textbf{HB} & \textbf{GreedyH} & \textbf{\sys} \\\hline
\multirow{3}{1cm}{{All Range}}      & {128}    & 1.38         & 1.85             & 1.38        & \textbf{1.16}             & \textbf{\emph{1.00}}         \\
                                         & {1024}   & 2.36        & 1.83             & \textbf{1.16}        & 1.33             & \textbf{\emph{1.00}}         \\
                                         & {8192}   & 4.51    & 1.79             & \textbf{1.12}        & 1.67             & \textbf{\emph{1.00}}         \\\hline
\multirow{3}{1cm}{{Prefix}}         & {128}    & 1.80         & 1.78             & 1.80        & \textbf{1.20}             & \textbf{\emph{1.00}}         \\
                                         & {1024}   & 3.34        & 1.80             & \textbf{1.34}        & 1.49             & \textbf{\emph{1.00}}         \\
                                         & {8192}   & 6.40        & 1.70             & \textbf{1.20}        & 2.09             & \textbf{\emph{1.00}}         \\\hline
\multirow{3}{1cm}{{Permuted Range}} & {128}    & 1.38        & 4.67             & 1.38        & \textbf{1.35}             & \textbf{\emph{1.00}}         \\
                                         & {1024}   & 2.36       & 10.57            & 3.35        & \textbf{2.16}             & \textbf{\emph{1.00}}         \\
                                         & {8192}   & 4.52       & 25.85            & 9.34        & \textbf{3.82}             & \textbf{\emph{1.00}}
\end{tabular}
}
}%
\hfill
\subcaptionbox{\label{table:2drange} 2D workloads}{
\resizebox{.48\textwidth}{!}{
\begin{tabular}{cc|SSSS|S}
\textbf{Workload}                   & \textbf{Domain}      & \textbf{Identity} & \textbf{Wavelet}    & \textbf{HB} & \textbf{QuadTree} & \textbf{\sys} \\
\multirow{3}{*}{$\P \otimes \P$}                                   & {64 x 64}     & 2.35              & 3.40  & \textbf{1.41}     & 1.72     & \textbf{\emph{1.00}} \\
                                    & {256 x 256}   & 4.75              & 3.14  & 2.03     & \textbf{1.95}     & \textbf{\emph{1.00}} \\
                                    & {1024 x 1024} & 11.17             & 3.25  & 2.96     & \textbf{2.49}     & \textbf{\emph{1.00}} \\\hline
\multirow{3}{*}{$ \R \otimes \R$}                                    & {64 x 64}     & 1.54              & 3.59  & \textbf{1.45}     & 1.72     & \textbf{\emph{1.00}}  \\
                                    & {256 x 256}   & 2.64              & 3.37  & 1.91     & \textbf{1.79}     &  \textbf{\emph{1.00}} \\
                                    & {1024 x 1024} & 5.57              & 3.34  & 2.54     & \textbf{2.09}     & \textbf{\emph{1.00}} \\\hline
\multirow{3}{*}{$ \begin{bmatrix} \R \otimes \T \\ \T \otimes \R \end{bmatrix} $}                                    & {64 x 64}     & 5.00              & 7.00  & \textbf{3.51}     & 4.13     & \textbf{\emph{1.00}} \\
                                    & {256 x 256}   & 13.68             & 8.52  & 7.88     & \textbf{6.69}     & \textbf{\emph{1.00}} \\
                                    & {1024 x 1024} & 38.84             & \textbf{10.31} & 13.91    & 10.49    &  \textbf{\emph{1.00}} \\\hline
\multirow{3}{*}{$\begin{bmatrix} \P \otimes \I \\ \I \otimes \P \end{bmatrix}$}
                                   & {64 x 64}     & \textbf{1.11}              & 5.26  & 2.08     & 3.32     & \textbf{\emph{1.00}} \\
                                    & {256 x 256}   & \textbf{1.44}              & 6.11  & 4.05     & 4.71     & \textbf{\emph{1.00}} \\
                                    & {1024 x 1024} & \textbf{1.99}              & 6.79  & 7.27     & 6.81     & \textbf{\emph{1.00}} \\
\end{tabular}
} } \vspace{-1ex}
\caption{\label{tab:lowd} Error, measured as \errRatio, of (a) four competing methods evaluated on three 1D workloads for varying domain sizes, and (b) of four competing methods on a variety of 2D workloads. (Best competitor in {\bf bold}.)}
\end{table*}

\subsection{Measurement+Reconstruction}

The algorithm described in \cref{sec:running} for computing $ \AA \x $ without materializing $\AA$ is shown formally in \cref{alg:fastikron}.  This algorithm is used for both measurement and reconstruction of strategies produced by $\optk$, $\optkk$, and $\optm$.  
%
%
%

\vspace{-1em}
\begin{algorithm}
\caption{Kronecker Matrix-Vector Product} \label{alg:fastikron}
{\small
\begin{algorithmic}[1]
\Procedure{kmatvec}{$A_1, \dots, A_d, x$}
\State $m_i, n_i = $ \Call{shape}{$A_i$}
\State $N_d = \prod_{i=1}^d n_i$
\State $Y_d = x$
\For{$i = d, \dots, 1$}
\State $Z_i = $ \Call{transpose}{\Call{reshape}{$Y_i$, $N_i/n_i$, $n_i$}}
\State $Y_{i-1} = A_i Z_i$
\State $N_{i-1} = N_i * m_i / n_i$
\EndFor
\State \Return \Call{flat}{$Y_0$}
\EndProcedure
\end{algorithmic}}
\end{algorithm}
\vspace{-1.5em}

\section{Additional experiments}

In this section we look deeper into the utility of HDMM compared to competing algorithms from the select-measure-reconstruct paradigm.  We also show that HDMM can improve DAWA \cite{li2014data}, which is a state-of-the-art algorithm outside of this paradigm.  

\subsection{One- and two-dimensional workloads}

The accuracy of \sys on low dimensional workloads is important because $\optk$ decomposes the optimization of high-dimensional workloads into single-dimensional sub-problems.  In addition, \sys can be used as a replacement for existing methods in low dimensions.   In this section we first compare \sys with a variety of data-independent algorithms, all of which fall into the select-measure-reconstruct paradigm. 
We evaluate \sys on range query workloads, comparing against existing algorithms designed specifically for such workloads: these are
Wavelet~\cite{xiao2011differential},
HB~\cite{qardaji2013differentially},
GreedyH~\cite{li2014data} (1D only),
QuadTree~\cite{cormode2012differentially} (2D only).

\cref{tbl:marg-1d} shows the results for 1D workloads, for which \sys has the lowest expected error in all cases.  The margins of improvement are sometimes modest, we believe because, for 1D workloads, competing approaches have found close-to-optimal solutions.  
The central assumption made by these methods is that the workload queries tend to exhibit locality, in that nearby elements of the domain are typically queried for together.  No such assumption is made by \sys, and the third workload (Permuted Range) highlights this: \sys is the only method that offers acceptable utility.  

For 2D workloads, \cref{table:2drange} shows that \sys outperforms all data-independent competitors on various workloads composed from Range, Prefix and unions thereof. In addition, the error improvements offered by \sys are more substantial (as much as 10$\times$), suggesting that they grow as the number of dimensions increases.

Importantly, without \sys, to achieve the best error rates for these 1D and 2D tasks, one has to choose between many algorithms (Wavelet, HB, GreedyH, Quadtree) each of which is best for some workload; \sys can replace them all and improves error uniformly.

\subsection{Marginals workloads}

We now evaluate \sys on $8$-dimensional data where each attribute domain has size 10, so that $N=10^8$.  Workloads are defined by $K$ where for a given $K$, the workload includes all $i$-way marginals where $i \leq K$. We compare \sys with three techniques: \Identity, \LMW and DataCube~\cite{ding2011differentially}.

As shown in \cref{tab:highd_marginals}, \sys outperforms the baselines in all target settings; the magnitude of the improvement depends on $K$.  \LMW is nearly optimal for small $K$, whereas \Identity is nearly optimal for large $K$, but \sys improves on both substantially for $K=3,4,5$.  For each experimental setting, one or the other baseline provides low error. But we emphasize the value of fully automated optimization: the algorithm designer is not forced to select the appropriate baseline, which will vary with domain size and workload.  
\begin{table}[h!]
\centering
\resizebox{.4\textwidth}{!}{
\begin{tabular}{r|SSS|c}
\textbf{Workload} & \multicolumn{1}{c}{\textbf{\Identity}} & \multicolumn{1}{c}{\textbf{\LMW}} & \multicolumn{1}{c}{\textbf{\DataCube}} & \textbf{\sys} \\\hline
$K=1$ & 435.19            & 1.18              & \textbf{1.12} & \textbf{\emph{1.00}}         \\
{2} & 43.89             & 1.43              & \textbf{1.03} & \textbf{\emph{1.00}}         \\
{3} & 8.37              & 1.96              & \textbf{1.15} & \textbf{\emph{1.00}}         \\
{4} & 2.73              & 3.03              & \textbf{1.21} & \textbf{\emph{1.00}}         \\
{5} & \textbf{1.33}              & 4.95              & 1.36 & \textbf{\emph{1.00}}         \\
{6} & \textbf{\emph{1.00}}              & 9.21              & 1.67 & \textbf{\emph{1.00}}         \\
{7} & \textbf{1.07}              & 18.21             & 2.99 & \textbf{\emph{1.00}}         \\
{8} & \textbf{1.06}              & 24.94             & 5.76 & \textbf{\emph{1.00}}
\end{tabular}} 
\caption{\label{tab:highd_marginals} Error, measured as \errRatio, on workloads of all up-to-$K$-way marginals on domain size of $10^8$.}
\end{table}

\clearpage

\subsection{Improving DAWA}

In the empirical study performed by Hay et al. \cite{hay2016principled}, the DAWA algorithm~\cite{li2014data} was one of the best performing algorithms for 1D and 2D linear query workloads.  The algorithm is data-dependent, using part of the privacy budget in a first stage that finds a partition of the data into contiguous regions that are well-approximated by uniformity. The second stage of the algorithm, is an instance of the select-measure-reconstruct pattern which uses GreedyH.  To show the value of \sys in improving the state-of-the-art, we modify DAWA by replacing GreedyH with \sys and measured the impact.  Since DAWA is data-dependent, we use a range of datasets and dataset sizes taken from \cite{hay2016principled} for the evaluation.  The workload is Prefix, which is a workload that GreedyH was designed to support.

To evaluate the impact, we measure the ratio of error between the modified algorithm and the original DAWA.  \cref{table:dawa} reports the min, median, and max improvement to GreedyH across the 5 datasets.  Maximum error improvements approach a factor of two in many cases; an impressive result considering that DAWA has been carefully tuned and outperforms most other algorithms in the literature.
\begin{table}[h!]
\centering
\resizebox{0.8\columnwidth}{!}{
\begin{tabular}{c|ccc|ccc|}
\multirow{2}{1.2cm}{{\bf domain \\ size}}&   \multicolumn{3}{c|}{{\bf data size} = 1000} & \multicolumn{3}{c|}{{\bf data size} = 10000000} \\
 & \textbf{min} & \textbf{median} & \textbf{max} & \textbf{min} & \textbf{median} & \textbf{max}\\\hline
 256             & 1.04         & 1.12            & 1.7          & 1.18         & 1.25            & 1.44         \\
 1024            & 1.04         & 1.15            & 1.91         & 1.15         & 1.37            & 1.92         \\
 4096            & 1.08         & 1.20             & 1.84         & 1.45         & 1.80             & 2.28 \\ \hline
\end{tabular} }
\caption{\label{table:dawa} Error ratio between modified DAWA and original DAWA. Min/median/max error across 5 datasets (Hepth, Medcost, Nettrace, Patent, Searchlogs~\cite{hay2016principled}) conforming to three domain sizes and two data sizes ($\epsilon=\sqrt{2}$).}
\end{table}

\section{Evaluating internals of HDMM}

\subsection{Hyper-parameter p}

\begin{figure}[h!]
\centering
\includegraphics[width=0.8 \linewidth]{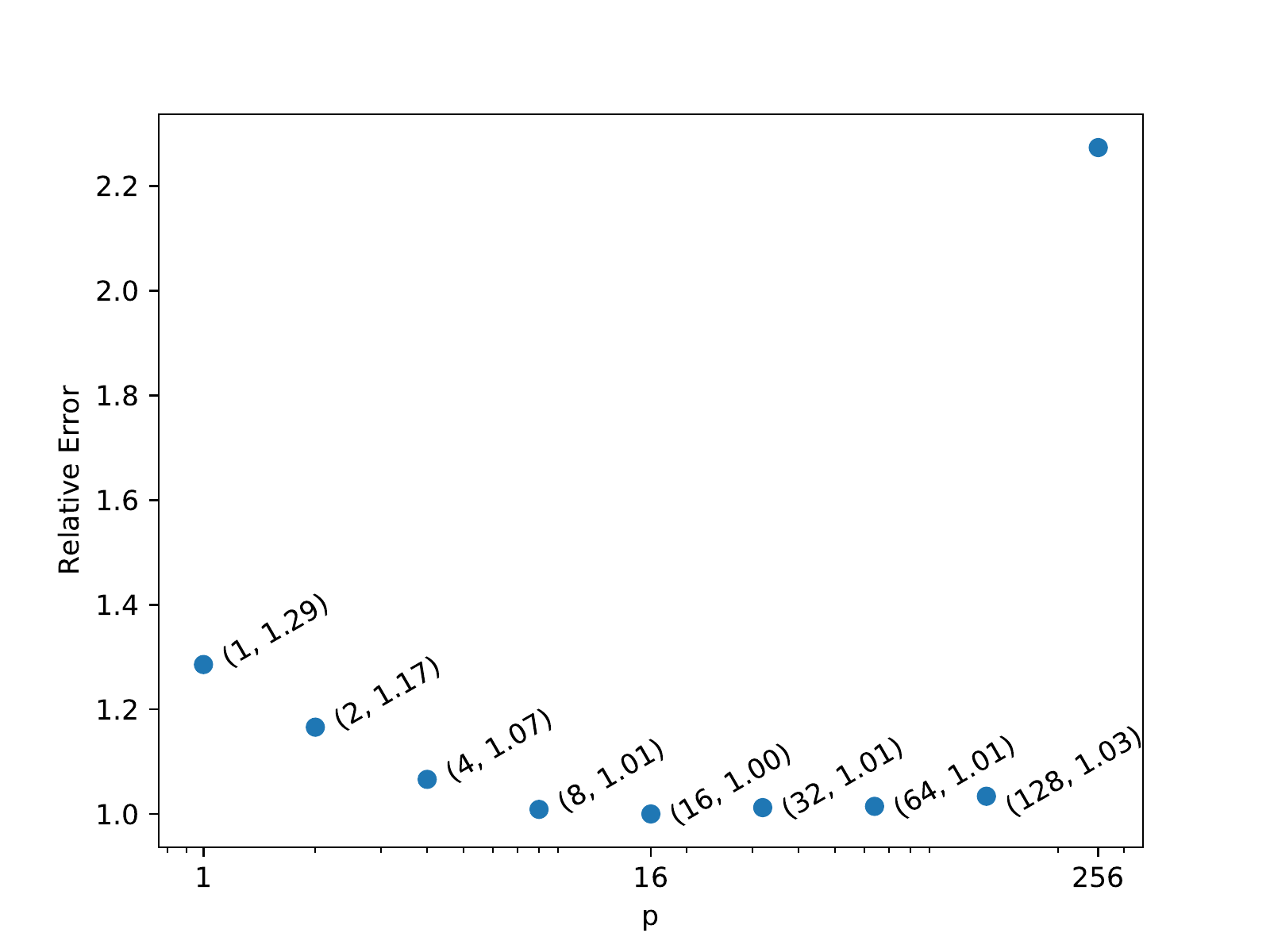}
\caption{Error of $\optgp$ for different settings of $p$.}\label{fig:hyperparam}
\end{figure}

\cref{fig:hyperparam} shows the error of $\optgp$ as function of $p$ on the workload of all range queries, for a domain of size $256$.  The error is approximately the same for any setting of $p$ between $8$ and $128$.  For $p \leq 4$, the search space is not expressive enough, while for $p \geq 256$ the search space is too expressive, causing the optimization to find poor local minima. 
\\
\subsection{Distribution of Local Minima}

\begin{figure}[h!]
\centering
\includegraphics[width=0.8\linewidth]{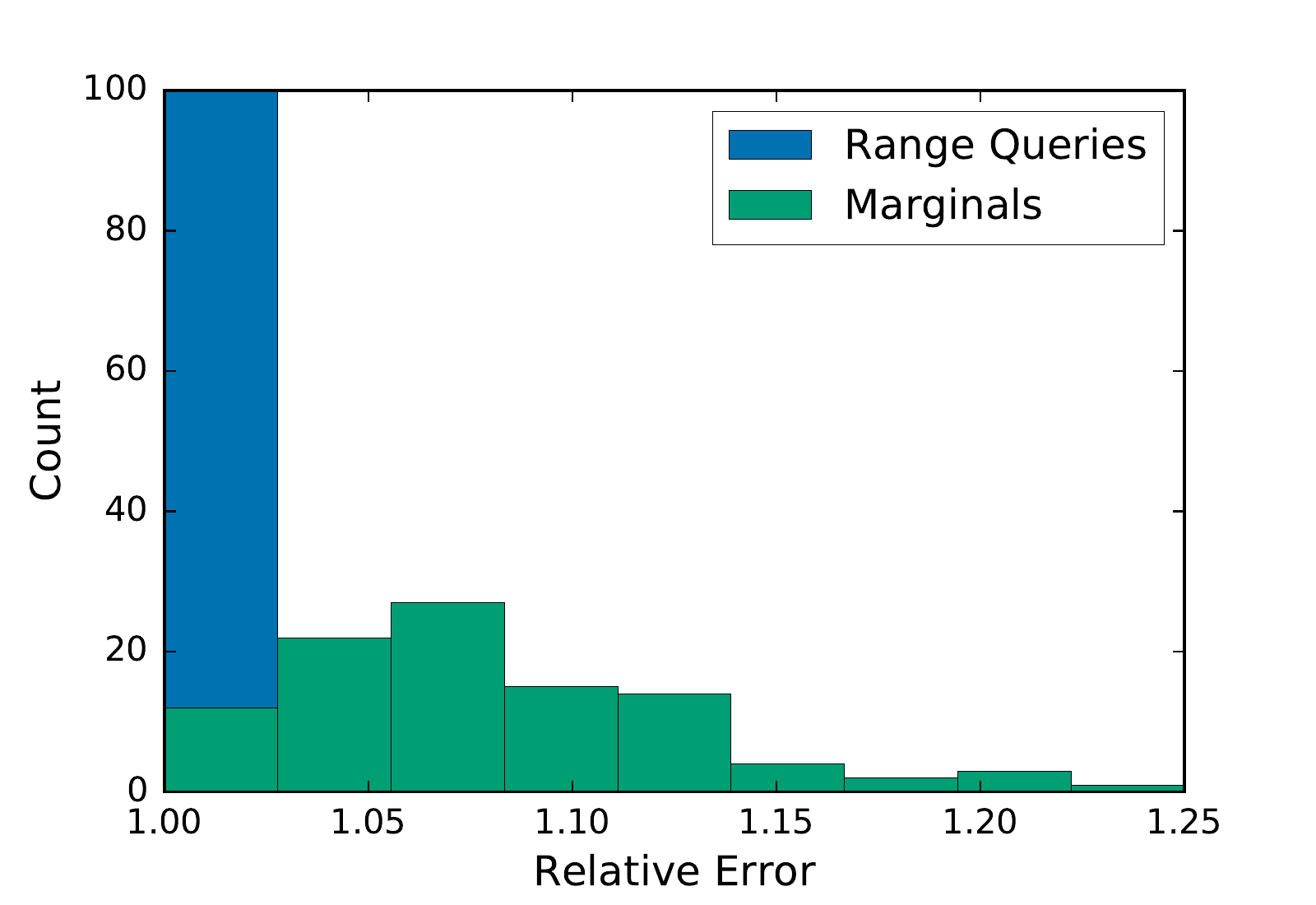}
\caption{Error distribution for locally optimal strategies.}\label{fig:errordist}
\end{figure}

Because the optimization problem is not convex, we are only able to find locally optimal strategies.  The quality of a locally optimal strategy depends on the initial guess.  We use random initialization, and show the distribution of local minima across $100$ random restarts for two workloads: range queries on a domain of size 256 (optimized with $\optgp$), and up-to-4-way marginals on a domain of size $10^8$ (optimized with $\optm$).  As shown in \cref{fig:errordist}, the distribution of local minima is very concentrated for the range query workload, indicating that no random restarts are necessary.  The distribution of local minima for the marginals workload varies more, but about 25\% of the local optima were within $1.05$ of the optimal strategy, so only a handful of restarts would be necessary in practice.

\subsection{Strategy visualization}

\begin{figure}[h!] 
\centering
\includegraphics[width=0.8 \linewidth]{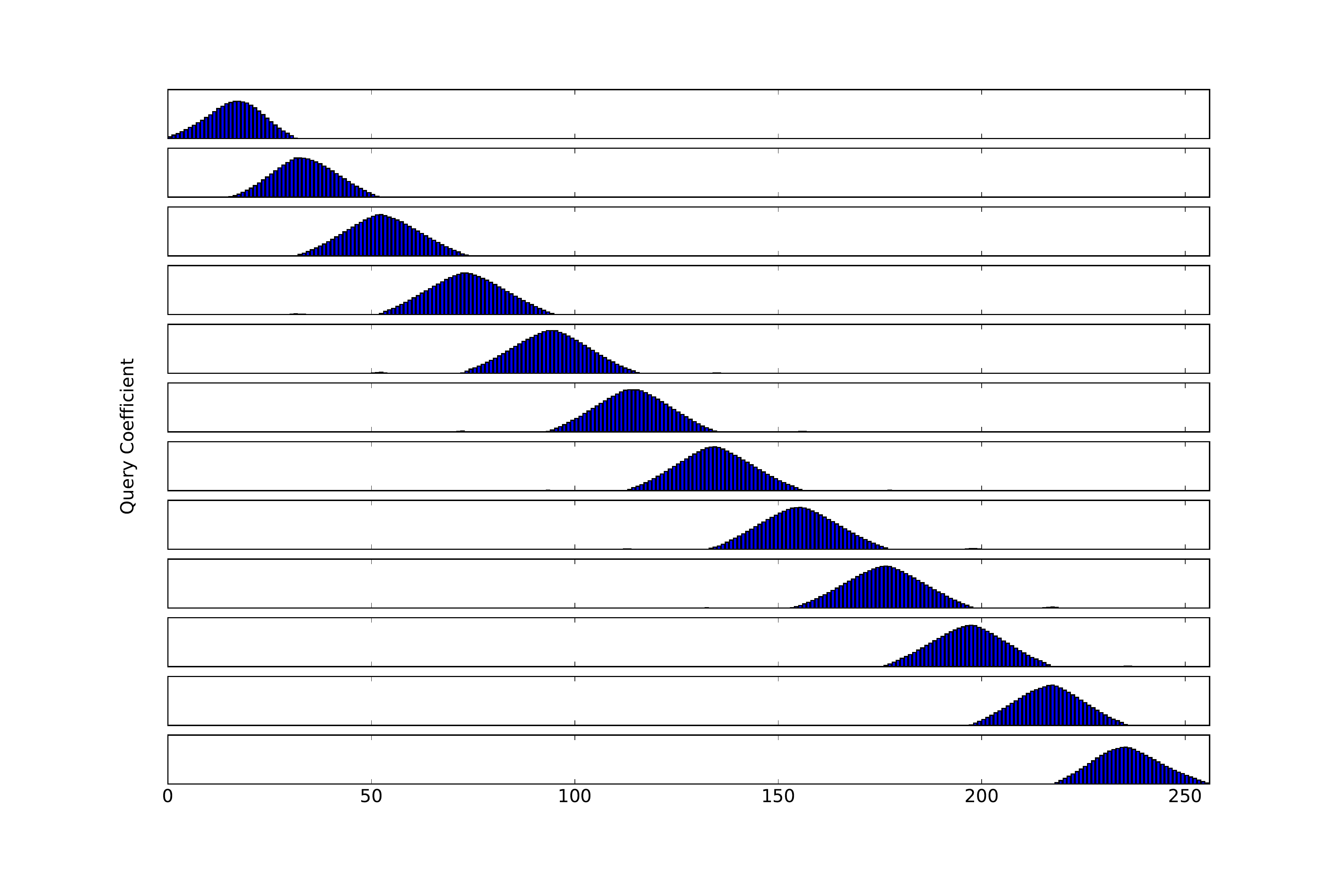}
\caption{Each of the 13 non-identity queries in the output of $\optgp$ for the workload of all range queries.}\label{fig:range_strategy}
\end{figure}

To provide some insight into the solutions found by $\optgp$ we show a visualization of the output of $\optgp$ in~\cref{fig:range_strategy}, which shows an understandable structure, but, interestingly, one that is not consistent with the hierarchical structures common to many heuristic strategy selection methods.  The visualization illustrates each of the rows of the $p=13$ non-identity queries in the output of $\optgp$, when optimizing the workload of all range queries on a domain of size 256.  The $x$-axis represents the cells in the data vector and the $y$-axis represents the weight on that cell in the query.  The identity queries are not plotted, but weights on those queries can be derived from the non-identity weights.  

\subsection{OPT$_0$ vs. OPT$_\otimes$}

\begin{figure}[h!]
\centering
\includegraphics[width=\linewidth]{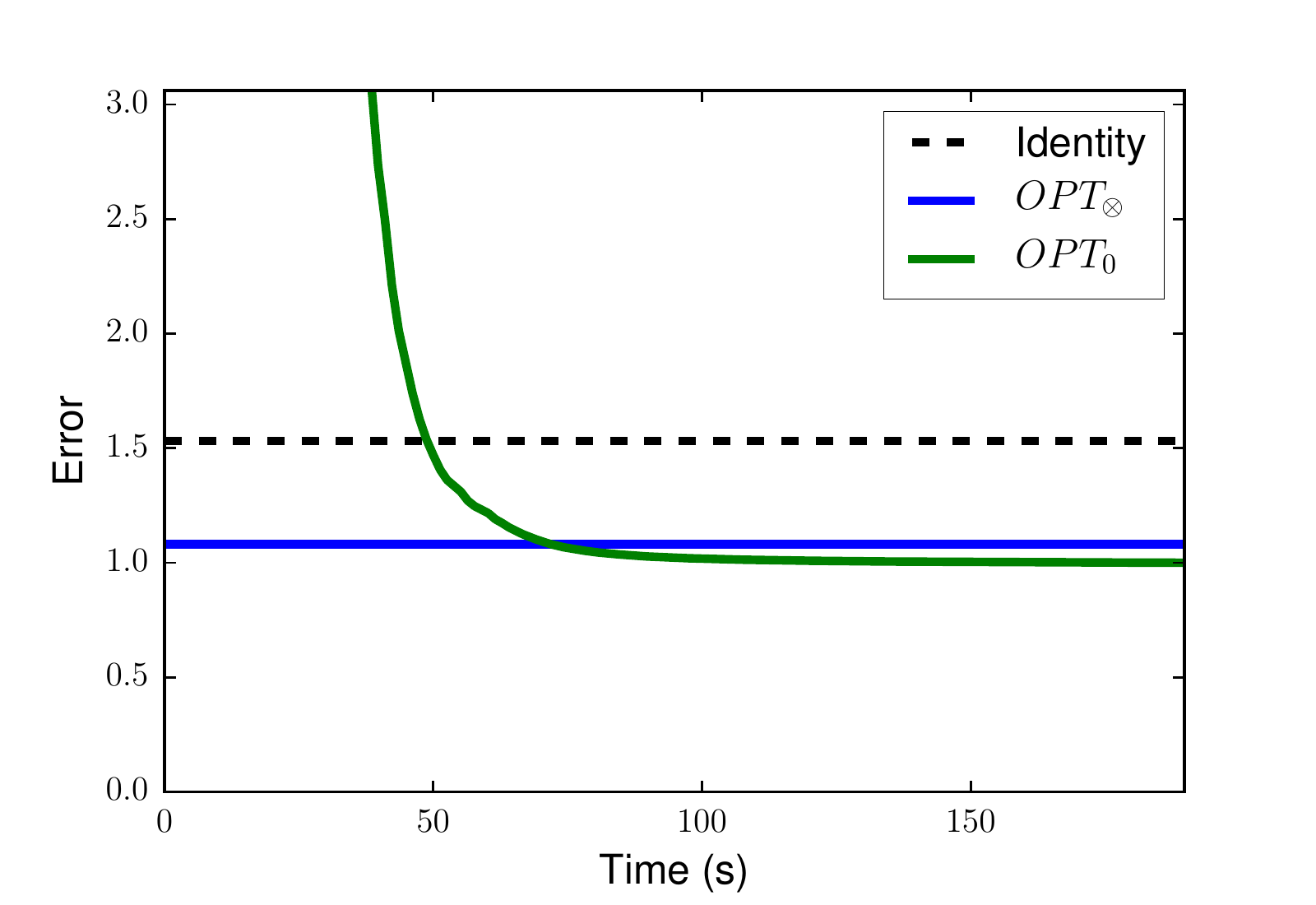}
\caption{Solution quality vs. time for $\optgp$ on all 2D range queries}\label{fig:kron_vs_general}
\end{figure}

We now compare $\optgp$ with $\optk$ on a workload where both methods are applicable and scalable: all 2D range queries over a $64 \times 64$ domain.  
\cref{fig:kron_vs_general} shows that $\optgp$ can find a slightly better strategy, since its search space is more expressive, but it takes much longer to converge.

\subsection{Scalability of optimization}

We now look deeper into the main optimization routines central to \sys.  In particular, we measure the scalability of $\optgp$ as a function of the domain size, and the scalability of $\optm$ as a function of the number of dimensions.  Since $\optk$ makes calls to $\optgp$ and $\optkk$ makes calls to $\optk$, the scalability of these optimization routines can be readily understood in terms of the scalability of $\optgp$

\begin{figure}[h!]
\centering
\includegraphics[width=0.48 \linewidth]{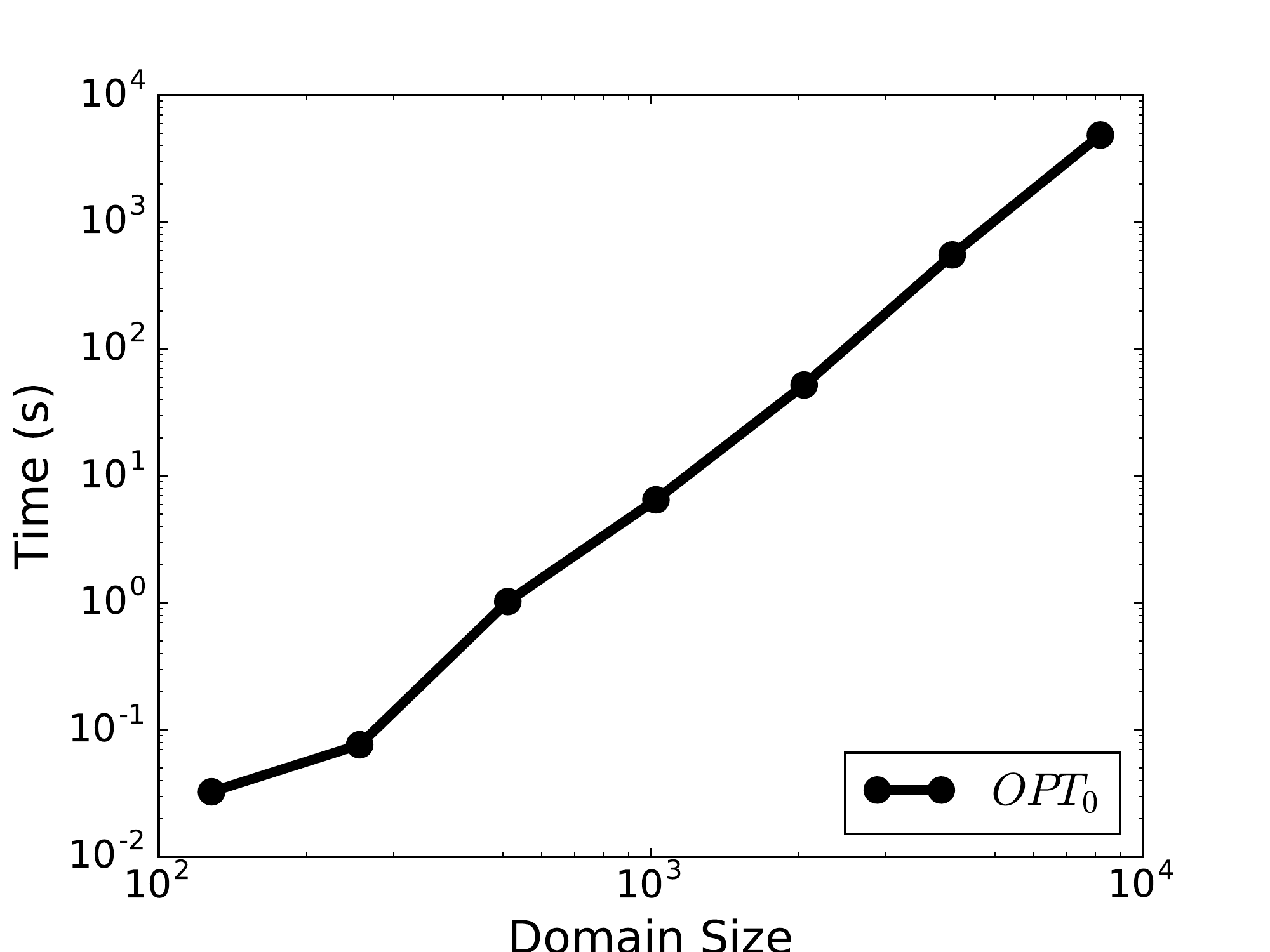}
\includegraphics[width=0.48 \linewidth]{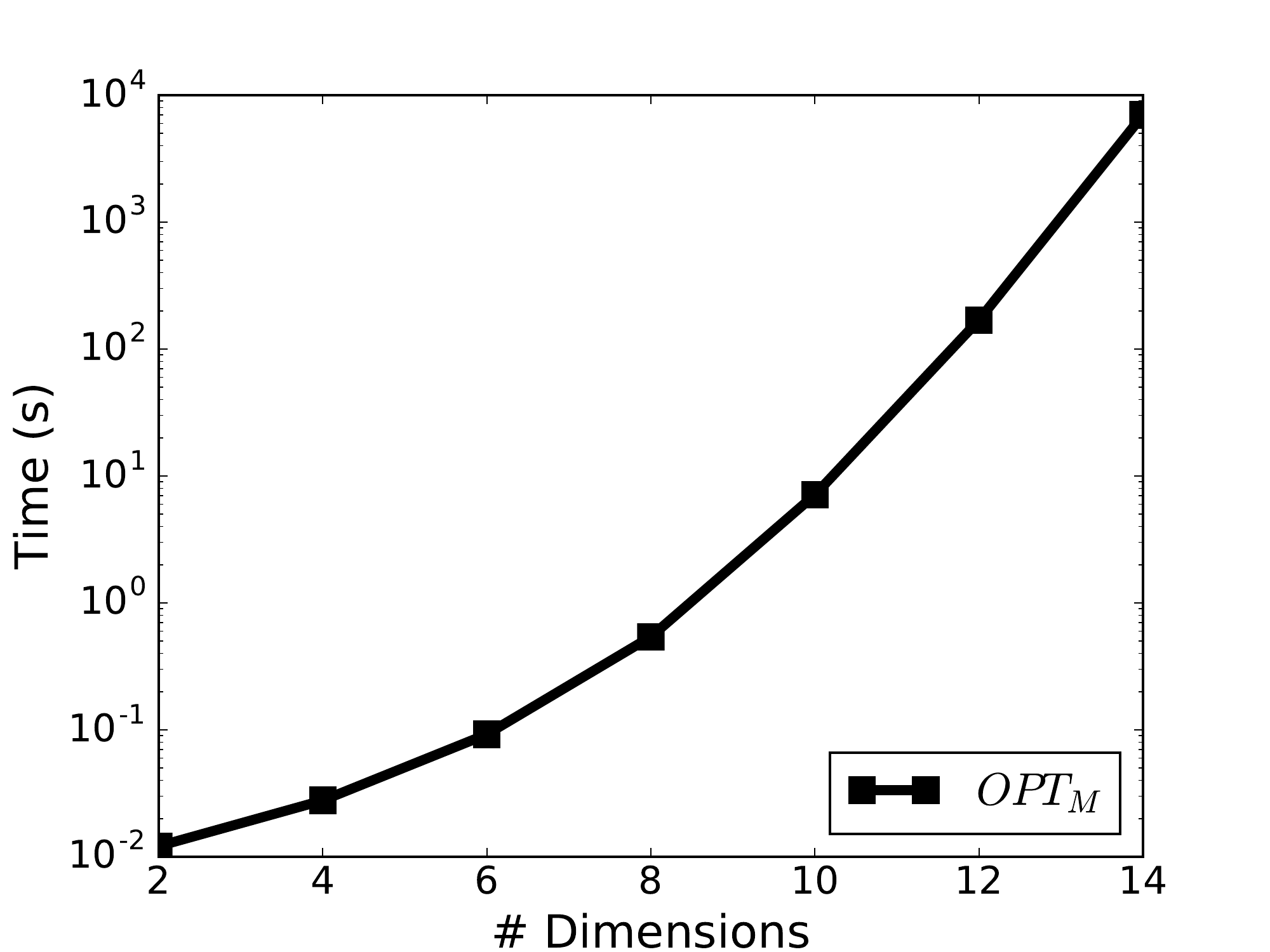}
\caption{Domain size $N$ vs. time for $\optgp$ (left) and number of dimensions $d$ vs. time for $\optm$ (right).} \label{fig:scalability-opt}
\end{figure}

\cref{fig:scalability-opt} shows that $\optgp$ scales up to domains as large as $N=8192$, and it runs in less than $10$ seconds for $N=1024$.  Thus, $\optk$ and $\optkk$ scale up to arbitrarily large domains, as long as the size of the largest attribute domain is no greater than $n_i=8192$.

\cref{fig:scalability-opt} also shows that $\optm$ scales up to $14$ dimensional domains, and up to $10$-dimensional domains in less than $10$ seconds.  Recall that the scalability of $\optm$ does not depend on total domain size.  Thus, it will take approximately the same amount of time for a binary domain of size $N=2^d$ as it would for any other $d$-dimensional domain, such as $ N = 10^d $. 

\end{document}